\RequirePackage{fix-cm}
\documentclass[smallextended]{svjour3}       
\smartqed  
\usepackage{graphicx}
\usepackage{amsmath}
\usepackage{amsfonts}
\usepackage{amssymb}
\usepackage{mathrsfs} 
\usepackage[toc]{appendix}
\usepackage{pst-node,pst-3d,pst-text,pst-tree}
\usepackage{graphicx}
\usepackage{color}
\usepackage{array}
\usepackage{tikz}
\usepackage{enumerate}
\usepackage{empheq}
\usepackage{float}
\definecolor{russiablue}{HTML}{116699}
\definecolor{russiabrown}{HTML}{CC6633}
\definecolor{russiagrey}{HTML}{CCCCCC}
\definecolor{indigo(dye)}{rgb}{0.0, 0.25, 0.42}

\usepackage{mathtools}
\usetikzlibrary{matrix,arrows}
\usepackage{lipsum}
\usepackage{blindtext}

\usepackage{verbatim}

\usepackage{wrapfig}
\usepackage{comment}

\usepackage{url}

\begin{document}

\title{The ancient Operational Code is embedded in the amino acid substitution matrix and aaRS phylogenies \thanks{This research was supported by Australian Research Council (ARC) Discovery Grant DP150100088 to Barbara R. Holland and Jeremy G. Sumner and Research Training Program scholarship to Julia A. Shore.}
}

\titlerunning{The ancient Operational Code is embedded}        

\author{Julia A. Shore$^1$       \and
        Barbara R. Holland$^1$ \and
        Jeremy G. Sumner$^1$ \and
        Kay Nieselt$^2$ \and
        Peter R. Wills$^3$
}


\institute{J. Shore \at
              \email{julia.shore@utas.edu.au}           
           \and            
             $^1$ School of Natural Sciences, University of Tasmania, Churchill Avenue, Tasmania, Australia 7005 \\
             $^2$ Integrative Transcriptomics, Interfaculty Institute for Biomedical Informatics, University of T{\"u}bingen, Sand 14, 72076 T{\"u}bingen, Germany \\
             $^3$Department of Physics and Te Ao Marama Centre for Fundamental Inquiry, University of Auckland, PB 92109, Auckland, 1142, Aotearoa, New Zealand \\
}

\date{Received: date / Accepted: date}

\maketitle

\begin{abstract}
The underlying structure of the canonical amino acid substitution matrix (aaSM) is examined by considering stepwise improvements in the differential recognition of amino acids according to their chemical properties during the branching history of the two aminoacyl-tRNA synthetase (aaRS) superfamilies. The evolutionary expansion of the genetic code is described by a simple parameterization of the aaSM, in which  (i) the number of distinguishable amino acid types, (ii) the matrix dimension, and (iii) the number of parameters, each increases by one for each bifurcation in an aaRS phylogeny. Parameterized matrices corresponding to trees in which the size of an amino acid sidechain is the only discernible property behind its categorization as a substrate, exclusively for a Class I or II aaRS, provide a significantly better fit to empirically determined aaSM than trees with random bifurcation patterns. A second split between polar and nonpolar amino acids in each Class effects a vastly greater further improvement.  The earliest Class-separated epochs in the phylogenies of the aaRS reflect these enzymes' capability to distinguish tRNAs through the recognition of acceptor stem identity elements via the minor (Class I) and major (Class II) helical grooves, which is how the ancient Operational Code functioned.  The advent of tRNA recognition using the anticodon loop supports the evolution of the optimal map of amino acid chemistry found in the later Genetic Code, an essentially digital categorization, in which polarity is the major functional property, compensating for the unrefined, haphazard differentiation of amino acids achieved by the Operational Code.
\end{abstract}

\section{Introduction}
\label{intro}
The extraordinary structural and functional specificity of cellular proteins depends on the exquisitely coordinated three-dimensional positioning of amino acid sidechain residues along their mobile peptide-bond backbones.  Two main processes are at play in producing functional proteins. The first is the sequence in which amino acids are concatenated in accordance with the genetic information supplied in the form of a messenger RNA and the rules of the genetic code used for protein synthesis. The second process is the folding of the protein backbone, which is controlled both thermodynamically and kinetically.  Within those constraints and the further influence of chaperones and other effectors, protein folding is not generally ``micro-managed'' by specific cellular components and energy-expending processes, but protein synthesis is.

The apparatus of protein synthesis is an enormous, intricately coordinated set of molecular machines. Specific enzymes catalyse reactions that produce individual amino acids from other metabolites. Aminoacyl-tRNA synthetase (aaRS) enzymes attach amino acids to their cognate tRNA adaptors and ribosomes join amino acids together to form proteins with mRNA-dictated sequences.  All of these processes are strictly controlled and the specificity of each processive step is maintained through the selective expenditure of thermodynamic free energy, available largely in the form of the diphosphate bonds that hold ATP together.

 Herein lies the central paradox of the origin of life, a variant of the ``chicken-egg problem''.  Specific protein catalysts cannot be produced in a cell unless the aaRS enzymes, themselves specific protein catalysts, are already present.  It is the aaRS that actually execute the rules of the genetic code, but they must first be produced by executing those rules.  Every cell from every organism from every taxon across the entire tree of life typically contains a suite of 20 aaRS, one for each canonical amino acid.  Evolution could not have produced them all at once, so what were the much simpler species from which the aaRS evolved?

 \subsection{Aminoacyl-tRNA synthetases}
 Each aminoacyl-tRNA synthetase (aaRS) enzyme attaches its amino acid substrate to the CCA tail on its cognate tRNA acceptor stem, maximally remote from the tRNA's anticodon.  It transpires that much of the information used by aaRS enzymes to determine the identity of tRNAs comprises structural features of the acceptor stem rather than the anticodon, which determines the specificity of the ribosomal process through its binding to a mRNA codon \cite{carter2018hierarchical}.  Indeed, these enzymes discriminate quite well between tRNA-like ``mini-helices'' comprising not much more than an acceptor stem with its CCA tail, attaching the correct amino acid with reasonable efficiency \cite{schimmel1993operational}. Furthermore, truncated aaRS ``urzymes'', completely lacking their anticodon recognition domains, retain significant catalytic activity, as well as specificity, for aminoacylation of an appropriate tRNA substrate \cite{li2013aminoacylating,CarterLife}.  Taken together, these findings attest to the existence of an ``operational code'' embedded in the interaction between the tRNA-acceptor stem and the aaRS catalytic unit.  Complementary tRNA-aaRS surface matching, involving acceptor stem tRNA identity elements and aaRS amino acid sidechain residues close to the catalytic site, appears to have predated the advent of either the tRNA anticodon loop or aaRS anticodon-recognition domains \cite{schimmel1993operational}.
 
 In every contemporary cell the aaRS enzymes are split between two superfamilies, labelled Class I and Class II.  The split is even, with enzymes for the same 10 amino acids found in the same Class across the entire tree of life, except that in Archaea the lysyl-RS is a Class I instead of Class II-type enzyme. Each aaRS superfamily has an origin and phylogeny which apparently predates the Last Universal Common Ancestor (LUCA) of all organisms \cite{o2003evolution,wolf2007origin,caetano2013structural} or at least the transition across the ``Darwinian threshold'' and the slowing down of horizontal gene transfer \cite{vetsigian2006collective}.  A constellation of biochemical, bioinformatic and theoretical studies indicate that the dual phylogenies of the two aaRS superfamilies are most probably a remnant of the expansion of the universal genetic code from early, simpler beginnings rather than their being an artefact proteins taking over a ribozymally-operated genetic code in a pre-existing RNA-World \cite{carter2017interdependence,wills2018insuperable}.

 The code appears to have started as a very crude mapping between two distinguishable pools or sets of amino acids and two \emph{operationally} differentiable\footnote{We use``differentiable'' rather than ``distinguishable'' when we want to emphasise the means of making a distinction, rather than simply the existence of a distinction, especially among amino acids.} sets of primitive codons (implicitly nucleotide triplets) \cite{carter2017interdependence}.  That is to say, the aaRS Class separation is a palimpsest of an ancient binary code based on just two \emph{operationally} distinguishable enzymatic activities with crudely separable amino acid-to-codon assignment specificities: amino acids \textit{a}  mapped onto codons \textit{A} by an ancestral Class I aaRS; and amino acids \textit{b} onto codons \textit{B} by an ancestral Class II aaRS.  The current study was conceived as a quest to find remnants of this primordial operational code in the link between aaRS phylogenies and the structure of amino acid substitution matrices.

\subsection{Amino acid substitution matrices}

Protein structure and function are relatively robust to sequence changes on a very broad scale, as evidenced by the extreme sequence variation in conserved core structures of homologous aminoacyl-tRNA synthetase (aaRS) enzymes drawn from  remote branches of the tree of life \cite{o2003evolution}. Of course changing sidechains whose specific chemistry is essential for the protein's function, such as arginine residues involved in the binding of ATP into an active site \cite{kaiser2018backbone}, can have a very large effect on a relevant enzyme's catalytic capabilities, but single amino acid substitutions more often have very little effect; hence the maintenance of functional identity in spite of sequence divergence during evolution. Furthermore, redundancies, regularities and internal symmetries in the coding table (assignment of amino acids to trinucleotide codons) ensure that the effect of single nucleotide mutations in coding sequences (copying errors) is minimised.

These effects can be seen in relative magnitudes of the entries in amino acid substitution matrices (aaSM), which provide a measure of the frequency and, by implication, the relative impunity with which one amino acid is found to, or may, substitute for another in an idealized representative protein.  Empirical substitution matrices are determined by constructing multiple alignments of homologous proteins and measuring the frequencies with which different amino acids appear in equivalent sequence positions.  Similarly, various physical parameters characterising the molecular chemistry of amino acids, such as size, polarity or hydrophobicity, have been used to construct scales of similarity for amino acid pairs and then to calculate theoretical values of the general likelihood of one amino acid substituting for another in protein sequences \cite{Grantham,Niefind1991,Atchley2005,Yampolsky2005}.

In assigning values to entries in a substitution matrix it is common bioinformatic practice for each protein sequence position to be considered independent of all others, so all of the functional effects of correlations between amino acid occupancies at different sites are lost.  While this means that two-dimensional amino acid substitution matrices provide no more than an approximate ``mean field'' representation of the functional effects of single amino acid changes, empirically and theoretically derived substitution matrices have both proved to be remarkably useful for a range of bioinformatic purposes, especially protein sequence alignment and phylogenetic inference.

Pokarowski \emph{et al.}  \cite{pokarowski2007ideal} used a parametric approach to compare a large number of diverse amino acid substitution matrices and establish correlations between them.  It was found that three almost completely independent factors are the major determinants of amino acid substitutability: whether or not pairs of amino acids are (i) hydrophobic or polar, (ii) large or small, or (iii) occurring in or absent from peptide backbone loops.  Although this result may be unsurprising, given the evident architectural constraints and chemical differences between the interior and exterior of globular proteins, it draws emphatic attention to the kind of ``nano-sensing'' of amino acid sidechain variability that evolution had to achieve: in order to create functionally specific proteins it was necessary to attain some means of differentiating the chemical effects of diverse amino acid sidechains.

In a parallel study of regularities in the tRNA-mediated map from codons to amino acids, Carter and Wolfenden \cite{carter2015trna} found that the anticodon ``measures'' the polarity of a tRNA's cognate amino acid, whereas tRNA identity elements in the helical acceptor stem at the other end of the molecule, close to where the amino acid is attached, show an orthogonal correlation with amino acid size.  Thus, as has long been evident from the regularities that make it robust against errors \cite{haig1991quantitative,koonin2009origin},  the genetic code is ordered principally as a map of amino acid sidechain properties.  In this study we seek traces of the emergence of that order in the dual phylogenies of the aaRS enzymes.

\subsection{aaRS phylogenies}  
Di Giulio \cite{di2001origin} identified a problem that is particularly acute in studies of the deep phylogenies of the aminoacyl tRNA synthetase (aaRS) protein families.  There is circularity in the logic of using amino acid similarities implicit in substitution matrices to align proteins while simultaneously using such alignments to determine empirical substitution matrices.  Elsewhere, attempts have been made to minimise the effects of this problem by giving phylogenetic weight to information pertaining to structural and functional homology, rather than relying on apparent sequence homology \cite{o2003evolution,Wolf1999,SmithHartman2015,Unvert2017}.

We avoid this problem altogether by mapping the deep phylogeny of the aaRS families directly onto amino acid substitution matrices, without applying these matrices to the analysis of protein sequences. We explore the possibility that the genetic code evolved from a binary root in a much simpler world where self-sustaining binary coding was achieved by two distinguishable populations of aaRS-like ``assignment catalysts'', progenitors of current Class I and II enzymes.

This original binary ``operational code'' may have depended on no more than the distinction between (i) a primitive Class I aaRS, with a propensity for larger amino acid substrates \textit{a}, binding to the minor groove side of tRNA-like minihelices bearing a C, A or G base at position 2 of the acceptor stem, defining a codon class \textit{A} through pairing to the middle base of the triplet comprising the minihelix's presumptive ``ghost'' anticodon (tRNA positions 70--72) \cite{DiGiulio2004}, and (ii) a primitive Class II aaRS, with a propensity for smaller amino acids \textit{b}, binding to the major groove side of minihelices with a U, A or G base at position 2, i.e., \textit{B} codons \cite{carter2015trna,carter2018hierarchical}. Furthermore, a large body of evidence now supports the hypothesis \cite{RodinOhno1995} that the genes for the binary encoding of the primitive, ancestral Class I and II assignment catalysts were complementary strands of a single information-bearing nucleic acid molecule \cite{carter2017interdependence}.

\subsection{Origin of Classes I and II aaRS}
We do not wish to downplay the problem of how such an apparently simple binary system of assignment catalysts \{I:\textit{a}$\rightarrow$\textit{A}, II:\textit{b}$\rightarrow$\textit{B}\} could have emerged from an initial dynamic state in which nucleic acids and peptides were synthesised in an essentially unordered, albeit co-dependent, fashion.  Synthesis of the aaRS \{I, II\} pair requires a mutual reaction system that is intrinsically autocatalytic: because each is envisaged to incorporate both \textit{a} and \textit{b} amino acids, each catalyst requires its partner as well as itself for its own synthesis.  Such cooperation is notoriously hard to achieve between entities that rely on the same resources for their production: phenotypes must cluster according to their degree of genetic relatedness to rescue the system from disruption by non-contributing (parasitic) variants \cite{Hamilton}.  
What is more, any coding system based on separate assignment catalysts is necessarily ``reflexive'' in the sense that it must contain, and therefore first find, very rare genetic information \cite{wills1993self} which, when interpreted according to the ``average'' assignment rules executed by the extant protein population, produces proteins that execute just the rules required to produce themselves and no byproducts with the assignment activities {\textit{a}$\rightarrow$\textit{B}} and {\textit{b}$\rightarrow$\textit{A}} that would disrupt the code in question.

F{\"u}chslin and McCaskill \cite{fuchslin2001evolutionary} have shown that a cooperative RNA-peptide system able to execute the exclusive rules of a code can spontaneously emerge from an initial state containing polymers of both sorts (RNA and protein) with completely random polymer sequences, even when all possible assignments \{\textit{a}$\rightarrow$\textit{A}, \textit{a}$\rightarrow$\textit{B} \textit{b}$\rightarrow$\textit{A}, \textit{b}$\rightarrow$\textit{B}\} are initially equally likely.
The observed symmetry-breaking transition in the system dynamics that produces genetic coding has been described as ``quasi-species bifurcation'' \cite{wills2015emergence} because it involves the splitting of not only a distibution of ``statistical proteins'' \cite{woese1965evolution} but also a nucleic acid quasi-species \cite{eigen1971selforganization} into two much narrower codependent subtypes.  Such self-organisation can only occur in systems whose chemical processes are differentiated spatially as well as temporally \cite{fuchslin2001evolutionary}, and requires, at the very least, reaction-diffusion coupling \cite{Turing} to provide the necessary clustering tendency as described by Hamilton \cite{Hamilton}.

When RNA and protein sequences first started to become codependently ordered, there were likely fewer amino acids abundant in biochemical quantities than the twenty that are universally used for coded protein production.  On the other hand, there were in all likelihood quite a few that were available in what we would now consider trace quantities, but present at high enough concentrations to be significant players in any relevant reaction processes, especially peptide bond formation.  Thus the evolution of genetic coding specificity involved as much a narrowing down as an expansion of the repertoire for individual amino acid recognition.  

In this study we focus on the stepwise narrowing down of recognition specificity through a series of bifurcations from the initial division of all amino acids into two classes corresponding to aaRS recognition specificities as far as the current division into twenty exclusive molecular types; and we seek to align that process with the decomposition of the amino acid substitution matrix into a set of nested submatrices with an elementary, uniform structure.  We do this without any explicit reference to the genetic code's specificity in respect of the way amino acids are assigned to codons, thereby distinguishing our approach radically from that of Delarue \cite{delarue2007asymmetric}. 

We reemphasise that the entire evolution of aaRS enzymes with different specificities appears to have occurred before the advent of the hypothetical LUCA, or alternatively, in the epoch before the transition across the ``Darwinian transition'' \cite{vetsigian2006collective}.  During this very early phase of protein evolution, horizontal gene transfer is thought to have been so prevalent that genes were selected according to the advantage they conferred on large collectives of proto-organisms, not separately identifiable biological species.  And we must likewise emphasise that in any epoch in which there were only $n < 20$ distinct aaRS types, it must be assumed that there were also only $n$ distinguishable classes of amino acids, even though there may have been 20 or more chemically distinguishable amino acids incorporated into proteins \cite{wills2015emergence}.  Under such circumstances the product of a single gene was indeed a ``statistical protein'' \cite{woese1965evolution} comprising a distribution of molecules with individual sequences that may have differed very widely.

\subsection{Modelling aaRS phylogeny}
We begin by considering a model of code evolution based on bifurcations in the phylogentic tree of aaRS enzymes \cite{o2003evolution}, starting with $n = 2$, the ancestral Class I and II aaRS enzymes, and culminating with $n = 20$, the standard genetic code and canonical suite of aaRS enzymes.  In each epoch $n$ simply represents the number of distinguishable subsets of amino acids, irrespective of the actual number of chemical species, or their abundances, environmentally available.  For each epoch we consider an $n \times n$ aaSM for the functional exchangeability, within the extant population of ``statistical proteins'', of amino acids chosen from different distinguishable subsets of amino acids.  Thus, an epoch-specific aaRS type, itself comprising a statistical distribution of amino acid sequences, is considered to accept a recognisable subset of amino acids as equivalent substrates for attachment to a set of cognate proto-tRNAs.  
Because each increase in the aaSM dimension corresponds to a bifurcation in a branch of the aaRS phylogeny, any unique branching pattern (Fig. 1) corresponds to a unique sequence of matrices of dimension $2 \leq n \leq 20$, irrespective of branch lengths \cite{wills2015emergence}.    

Each expansion in the aaSM subsequent to the advent of the Class I and II aaRS progenitors requires the definition of a new parameter, so that the final $20 \times 20$ matrix can be limited to 19 free parameters compared with the 190 that populate empirical matrices.  Each bifurcation of substrate specificity in an aaRS phylogeny represents the expansion of the genetic code \cite{wills2015emergence} either (i) to accommodate assigned coding for a new amino acid; or (ii) to differentiate two subclasses of a class of amino acids defined by the substrate specificities of extant aaRS types (Fig. 1). For heuristic reasons, especially simplicity of nomenclature, we draw the Class I and II aaRS phylogenies as emanating from a hypothetical common root, which could be taken to represent disordered protein production \cite{WillsMutualOrdering}.  This convention has no effect on our analysis or conclusions.    

Perhaps the clearest real-life illustration that these two possibilities, amino acid addition or differentiation, are formally equivalent from the perspective of aaSM expansion is at the point of coalescence in the ancestry of tyrosyl- and tryptophanyl-tRNA synthetases.  Whether the common ancestor of these two enzymes consisted of a species for which only one of these two aromatic amino acids was a substrate, or alternatively, an enzyme that could not distinguish between tyrosine and tryptophan as substrates, is hard to determine \cite{fournier2015ancestral}.  Either way, at times prior to when tyrRS and trpRS could be identified as separate enzymes, entries in a putative aaSM accounting for tyr-trp swapping make no sense; in which case the rows and columns corresponding to other amino acids swapping with tyrosine and/or tryptophan should ``see'' that pair as a single amino acid ``type''.  These alternative but equivalent interpretations of aaRS phyologenetic trees are illustrated in Fig. 1.    

Some investigators \cite{hornoshornos,Bashford1998} have taken a group-theoretical approach to the description of code expansion in terms of sequential bifurcations, but apart from Delarue \cite{delarue2007asymmetric} they have not considered how the process was constrained by structural and functional properties of the aaRS enzymes.  What distinguishes our approach is that it makes no obligatory reference to the coding table that assigns nucleotide triplets to amino acids, relying solely on how evolving biochemistry could use differences in the properties of amino acids to create a map of those differences.  Such a process is necessarily ``reflexive'' in that the creation of the map relies on the effects of its use.

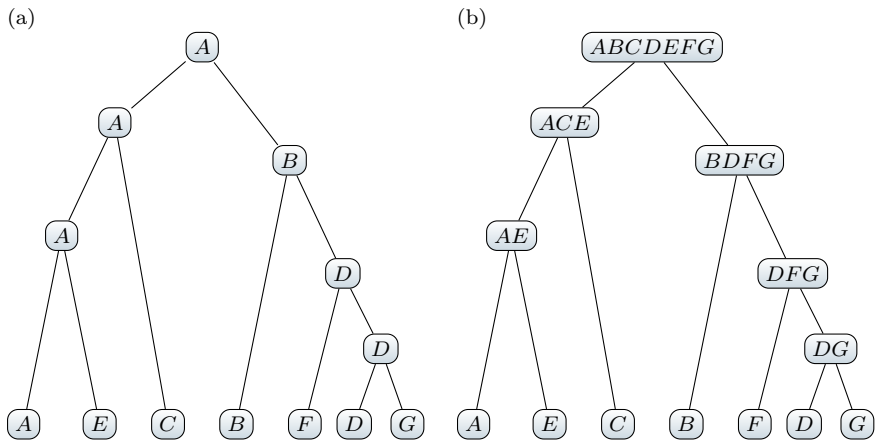
\begin{figure}
\begin{minipage}{0.49\textwidth}
(a) \\ \begin{tikzpicture}[level 1/.style={sibling distance=23mm},level 2/.style={sibling distance=14mm},level 3/.style={sibling distance=10mm},level 4/.style={sibling distance=7mm},
  every node/.style = {shape=rectangle, rounded corners,
    draw, align=center,
    top color=white, bottom color=indigo(dye)!20}]]
  \node {$A$}
    child[level distance = 1cm] { node {$A$}
    	child[level distance = 1.5cm] { node {$A$}
        	child[level distance = 2.5cm] {node {$A$} } 
            child[level distance = 2.5cm] {node {$E$} } }
        child[level distance = 4cm] { node {$C$} 
        	 } }
    child[level distance = 1.5cm]{ node {$B$} 
    	child[level distance = 3.5cm]{ node {$B$} }
        child[level distance = 1.5cm]{ node {$D$} 
        	child[level distance = 2cm]{node{$F$}} 
        	child[level distance = 1cm]{ node {$D$} 
            	child[level distance = 1cm] {node{$D$}}
                child[level distance = 1cm] {node{$G$}}  } 
        	}
          };
\end{tikzpicture} \end{minipage}
\begin{minipage}{0.49\textwidth}
(b) \\ \begin{tikzpicture}[level 1/.style={sibling distance=23mm},level 2/.style={sibling distance=14mm},level 3/.style={sibling distance=10mm},level 4/.style={sibling distance=7mm},
  every node/.style = {shape=rectangle, rounded corners,
    draw, align=center,
    top color=white, bottom color=indigo(dye)!20}]]
  \node {$ABCDEFG$}
    child[level distance = 1cm] { node {$ACE$}
    	child[level distance = 1.5cm] { node {$AE$}
        	child[level distance = 2.5cm] {node {$A$} } 
            child[level distance = 2.5cm] {node {$E$} } }
        child[level distance = 4cm] { node {$C$} 
        	 } }
    child[level distance = 1.5cm]{ node {$BDFG$} 
    	child[level distance = 3.5cm]{ node {$B$} }
        child[level distance = 1.5cm]{ node {$DFG$} 
        	child[level distance = 2cm]{node{$F$}} 
        	child[level distance = 1cm]{ node {$DG$} 
            	child[level distance = 1cm] {node{$D$}}
                child[level distance = 1cm] {node{$G$}}  } 
        	}
          };
\end{tikzpicture} \end{minipage} 

\caption{Alternative representations of aaRS phylogenetic trees in terms of different models of the evolution of genetic coding. Nodes represent aaRS enzymes and upper case letters represent the amino acid alphabetic specificity of the enzymes' catalytic activity (up to $n = 7$ letters).  In each case the root is artificial. (a) Amino acids are added to the coding alphabet one at a time as they become biochemically available. (b) All amino acids are available from the start but the substrate specificity of daughter enzymes becomes selective for newly distinguished subsets of the alphabet after each branch bifurcation.}
\end{figure}

\subsection{Overview}
Using the approach described above, we investigate how well the parametric structure of different aaSM structures, each corresponding to a particular phylogeny of aaRS types, can be quantitatively fitted to empirically derived aaSMs that are commonly used by bioinformaticians for the alignment of protein sequences and the construction of phylogentic trees. We find that trees which initially divide the amino acids into operational groups according to the Class I and II aaRS specificities  dictate 19-parameter fits that are better than randomly chosen branching patterns, on average. More refined groupings that divide amino acids into polar and nonpolar groups are even better.  The aaSM for a specific phylogeny derived from aaRS sequence information is among those of optimal construction.  The only aaRS phylogenies that give a better parameterized representation of empirical aaSMs than those based on the actual Class I:II distinction, are those that would have hypothetically arisen had the split of the initial operational code taken place on the basis of exclusive preferences for either polar or non-polar amino acids.  We interpret this result in light of the finding that the anticodon-based genetic code overwrote the operational code that depended largely on tRNA acceptor stem properties \cite{carter2015trna,carter2018hierarchical}. 

\section{Results}
\label{sec:1}

\subsection{Parameterized exchangeability matrices}
\begin{figure}

\begin{minipage}{0.15\textwidth}
\centering
\begin{tikzpicture}[level 1/.style={sibling distance=10mm},level 2/.style={sibling distance=25mm},
  every node/.style = {shape=rectangle, rounded corners,
    draw, align=center,
    top color=white, bottom color=indigo(dye)!20}]]
  \node {$ \alpha $}
    child[level distance = 4.5cm] { node {\em A} }
    child[level distance = 4.5cm]{ node {\em B}};
\end{tikzpicture}

\end{minipage}
\begin{minipage}{0.21\textwidth}
\centering
      \begin{tikzpicture}[level 1/.style={sibling distance=13mm},level 2/.style={sibling distance=10mm},
  every node/.style = {shape=rectangle, rounded corners,
    draw, align=center,
    top color=white, bottom color=indigo(dye)!20}]]
  \node {$ \alpha $}
    child[level distance = 2.5cm] { node {$ \beta $}
      child[level distance = 2cm]{node {$A$}}
      child[level distance = 2cm]{node {$C$}}}
    child[level distance = 4.5cm] { node {$B$} };
\end{tikzpicture}

\end{minipage}
\begin{minipage}{0.24\textwidth}
\centering
\begin{tikzpicture}[level 1/.style={sibling distance=15mm},level 2/.style={sibling distance=8mm},
  every node/.style = {shape=rectangle, rounded corners,
    draw, align=center,
    top color=white, bottom color=indigo(dye)!20}]]
  \node {$ \alpha $}
    child[level distance = 1.5cm] { node {$ \beta $}
      child[level distance = 3cm]{node {$A$}  }
      child[level distance = 3cm]{node {$C$} }}
    child[level distance = 2.5cm] { node {$ \gamma $}
      child[level distance = 2cm] { node {$B$}  }
      child[level distance = 2cm] { node {$D$} } };
\end{tikzpicture}
\end{minipage}
\begin{minipage}{0.37\textwidth}
\centering
\begin{tikzpicture}[level 1/.style={sibling distance=17mm},level 2/.style={sibling distance=10mm},level 3/.style={sibling distance=8mm},
  every node/.style = {shape=rectangle, rounded corners,
    draw, align=center,
    top color=white, bottom color=indigo(dye)!20}]]
  \node {$ \alpha $}
    child[level distance = 1cm] { node {$ \beta $}
      child[level distance = 2cm]{node {$ \delta $}
         child[level distance = 1.5cm] {node {$A$} }
         child[level distance = 1.5cm] {node {$E$} }  }
      child[level distance = 3.5cm]{node {$C$} }}
    child[level distance = 2cm] { node {$ \gamma $}
      child[level distance = 2.5cm] { node {$B$}  }
      child[level distance = 2.5cm] { node {$D$} } };
\end{tikzpicture}

\end{minipage}

\begin{minipage}{0.15\textwidth}
\[ \bordermatrix{ & A & B \cr
      A \phantom{i} & * & \alpha \cr
      B \phantom{i} & \alpha & *  \cr } \qquad \]
\end{minipage}
\begin{minipage}{0.21\textwidth}
\[ \bordermatrix{ & A & C & B \cr
      A \phantom{i} & * & \beta & \alpha  \cr
      C \phantom{i} & \beta & * & \alpha  \cr
      B \phantom{i} & \alpha &  \alpha & *  } \qquad \]
\end{minipage}
\begin{minipage}{0.24\textwidth}
\[ \bordermatrix{ & A & C & B & D \cr
      A \phantom{i} & * & \beta & \alpha & \alpha \cr
      C \phantom{i} & \beta  & * & \alpha & \alpha \cr
      B \phantom{i} & \alpha  &  \alpha & * & \gamma \cr
      D \phantom{i} & \alpha  & \alpha & \gamma & * } \qquad \]
\end{minipage}
\begin{minipage}{0.37\textwidth}  
\[  \bordermatrix{ & A & E & C & B & D \cr
      A \phantom{i} & * & \delta & \beta & \alpha & \alpha \cr
      E \phantom{i} & \delta & * & \beta & \alpha & \alpha \cr
      C \phantom{i} & \beta & \beta & * & \alpha & \alpha \cr
      B \phantom{i} & \alpha & \alpha &  \alpha & * & \gamma \cr
      D \phantom{i} & \alpha & \alpha & \alpha & \gamma & * } \qquad   \] 
\end{minipage}
\caption{Progressive parametric construction of an aaSM corresponding to an aaRS phylogeny.  Starting from a hypothetical common root, the first bifurcation gives two differentiable aaRS activities with specificities for amino acid subsets $A$ and $B$ and the exchangeability between $A$ and $B$ is $\alpha$.  Each subsequent bifurcation necessitates a new parameter, $\beta$ then $\gamma$ then $\delta$, for the exchangeability between the two amino acid subclasses newly differentiable by the daughter aaRS enzymes, one of which retains the label of the parent, as in Fig. 1.}
\label{treesMatrices}
\end{figure}
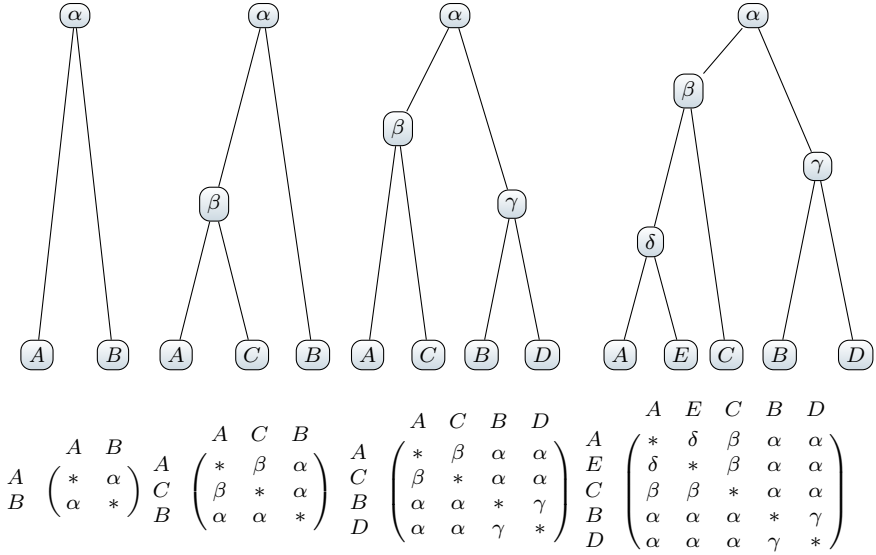

Given that the aaRS enzymes belonging to each of the two superfamilies can be traced to one of two progenitor proteins, either Class I or II, we envisage the genetic code as having first appeared as a minimal two-letter system  \cite{wills2015emergence}, with all possible amino acid substitutions being described in terms of a single exchangeability parameter $\alpha$ defining a symmetric $2\times2$ SM (Fig. \ref{treesMatrices}).   Subsequent to this first bifurcation, each further branch point in the phylogenetic tree represents the refinement of one of the Classes to expand its operational alphabet from $m$ to $m+1$ differentiable subsets of amino acid types; and $m+1$ subclasses of that aaRS Class.  This process is presumed to have been repeated until each canonical Class could recognise 10 distinct amino acids, \{arg, cys, gln, glu, ile, leu, met, trp, tyr, val\} for Class I and \{ala, asn, asp, gly his, lys, phe, pro, ser, thr\} for Class II.  As noted above, this analysis in no way prejudices the question as to whether alphabet expansion occurred as new substrates became available \cite{wong1975co,bernhardt2008evidence,ikehara2005possible} or extant indistinguishable substrates became distinguishable (Fig. 1).  

\newcommand{\+}[1]{\ensuremath{\mathbf{#1}}}
Use of an aaSM in phylogenetic analysis requires assumptions concerning the underlying evolutionary mechanism of amino acid substitution.  It is commonly assumed that substitution is a Markov process in which any two amino acids have the same propensity to replace one another and replacement occurs independently at all sites in a protein sequence.  In that case a symmetric ``exchangeability'' matrix $\+R$ can be constructed in which an entry $r_{i \leftrightarrow j}$ represents the relative propensity for amino acid $j$ to replace amino acid $i$ at any individual site (transition from state $i$ to state $j$), and vice versa.  The $r_{i \leftrightarrow j}$ values are generally obtained from estimates of the instantaneous rate (number of occurrences per unit time) $q_{ij}=\pi_j r_{i \leftrightarrow j}$ at which the change of state $i \rightarrow j$ is found to have occurred, averaged over all sites; $\pi_j$ is the fraction of sites in state $j$. An empirical estimate of $\+R$ is typically obtained from ``measurements'' of actual substitution rates $q_{ij}$ in a phylogeny reconstructed from a multiple sequence alignment of homologous proteins \cite{whelan2001general,dayhoff197822,le2008improved}.  

The formal properties of $\+R$ are such that if two states $a$ and $b$ are merged into a single identity $c$, reducing the number of distinguishable states from $n$ to $n-1$, then $r_{c \leftrightarrow i} = (\pi_a r_{a \leftrightarrow i} + \pi_a r_{b \leftrightarrow i})/\pi_c$, where $\pi_c = \pi_a + \pi_b$.  In the case that $r_{a \leftrightarrow i} = r_{b \leftrightarrow i}$ it is clear that both of these parameters are equal to $r_{c \leftrightarrow i}$.  In our analyses we consider the reverse process: the differentiation of a state $c$ into states $a$ and $b$.  We avail ourselves of the simplifying assumption $r_{a \leftrightarrow i} = r_{b \leftrightarrow i} = r_{c \leftrightarrow i}$, which means that the exchangeability $r_{c \leftrightarrow i}$ between any state $i$ and a parent state $c \neq i$ is conferred on both daughter states, $a$ and $b$.  In other words, the equivalence of $a$ and $b$ survives the differentiation of $c$ into $a$ and $b$, in particular, vis-\`a-vis exchange of $a$ or $b$ for $i $ ($ i \not\in \{a,b\} $), and vice versa. This assumption captures the idea that code expansion through aaRS evolution is conservatively progressive: any definitive ability of the extant aaRS population operationally to differentiate between subclasses of amino acids, starting with $n=2$ and culminating in $n=20$, is preserved unaltered from any evolutionary epoch into the next with the advent of newly defined $a \leftrightarrow b$ exchanges: the system ``learns'', in a rather slow and rigid manner, how to differentiate between amino acids.  While it is clear that the modern aaSM does not have the exactly parameterized form which this assumption imposes, it nevertheless serves as a reasonable {\em ansatz} upon which to explore the correlation between aaRS phylogeny and aaSM structure.

We represent any epoch in aaRS phylogeny as a rooted tree with $2 \leq n \leq 20$ leaves (taxa), the leaves corresponding to the distinguishable subclasses of amino acids at that stage of code evolution.  We label each vertex in a tree with a parameter that measures the exchangeability between members of the two amino acid subclasses immediately below that vertex, that is, the parameter labelling the vertex joining subclasses $a$ and $b$ is $r_{a \leftrightarrow b}$ (Fig. 2).  Therefore, for a binary tree of $n$ taxa, we assign only $ n-1 $ possible unique exchangeability values, with the label on the root of the tree, $\alpha$, representing the exchangeability between primordial Classes I and II aaRS.  Application of this nomenclature to the first stages of aaRS evolution is illustrated in Fig. 2, with the labels $A$ and $B$ representing the possible states following the initial bifurcation of aaRS activity into  Classes I and II.  Notice that when a subclass bifurcates, one of the daughter branches retains the original state index ($A$) and the other takes the next available letter in the alphabet ($C$), as in Fig. 1b, and the new parameter for exchangeability between the daughter states ($\beta$) overwrites the name of the parent state.

For two $ n \times n $ normalised exchangeability matrices \textbf{R} and \textbf{S} with entries $ r_{ij} $ and $ s_{ij} $ respectively, the residual sum of squares distance \textit{RSS} between \textbf{R} and \textbf{S} is defined as

\begin{equation}
 \textit{RSS} =  \sqrt{ \sum_{i=1}^{n} \sum_{j \not = i}^{n} (r_{ij} - s_{ij})^2}. 
\end{equation}

\noindent Typically, $\textbf{R}$ would be derived from an experimentally determined rate matrix such as WAG \cite{whelan2001general}, PAM \cite{dayhoff197822} or LG \cite{le2008improved} and $\textbf{S}$ would be obtained by fitting the $n-1$ parameters of an $ n \times n $ matrix of the form illustrated in Fig. \ref{treesMatrices}, noting that there is a 1:1 correspondence between the detailed form of such matrices and the hypothetical aaRS phylogenetic trees from which they are derived.  On the other hand, the final form of a parameterized matrix depends only on the overall topology of the tree from which it is derived, not on the temporal order in which the branching bifurcations take place.  That order does determine the form of the matrix relevant to an intermediate epoch $(2 < n < 20)$ and is therefore relevant to the use of such matrices for bioinformatic analysis, but it does not affect the goodness of fit of a parameterized $20 \times 20$ matrix $\textbf{S}$ to an empirical matrix $\textbf{R}$ according to the measure of Eq. (1).

We present results obtained by using the LG aaSM \cite{le2008improved} as a representative of empirically based exchangeability matrices. This particular aaSM was chosen as it is the most recently developed of widely used aaSMs and its underlying alignment data base is both extensive and reliable. We also conducted the same analysis using other aaSMs: PAM \cite{dayhoff197822}, BLOSUM \cite{henikoff1992amino}, JTT \cite{jones1992rapid} and WAG \cite{whelan2001general}. The results from these analyses were similar enough to the LG results that the same conclusions could be drawn.

\subsection{Types of parameterized exchangeability matrices}
Different types of twenty-taxon trees representing hypothetical aaRS phylogenies were constructed and their corresponding matrices were compared with an experimental matrix ($\textbf{R}$) using Eq. 1 as a goodness-of-fit criterion. For each type of tree, $N = 100,000$ random trees ($k=1:N$) of that type were generated, along with each tree's corresponding $20 \times 20$ parameterized exchangeability matrix ($\textbf{S}_k$) having the block structure and containing the $19$ free parameters as shown in Fig. 1. The free parameters of each ($\textbf{S}_k$) were determined by least squares fitting (Eq. (1)) to the empirical exchangeability matrix of choice. This procedure yielded $N$ distance measures for sample trees of any type. The distribution of distance measures for trees of different types were then determined.

Trees of the first type were labelled ``random'' and such trees were constructed by making random choices as to the order in which branches bifurcated.  Trees of the second type were labelled ``ten-ten'' and such trees were also constructed by making random branching decisions but in this case under the constraint that each branch of the initial binary split led to 10 leaves.  Trees of the third type were labelled ``I:II'' and such trees were ten-ten trees that were further constrained to ensure that the division between the two sets of 10 leaves corresponded to the amino acid specificities of the standard bacterial and eukaryotic Class I and II aaRS enzymes. Last, but not least, a single parameterized matrix of a fourth type was generated based on a phylogenetic tree ``pseq'' inferred from the amino acid sequences of actual aaRS proteins. The relevant tree is a preliminary result from the study of Popinga et al. \cite{Popinga}. 

Constructing the dual evolutionary trees for Class I and II aaRS proteins is a significant challenge, for the reasons previously outlined: it is to be expected that in the early stages of aaRS evolution there were fewer subsets of amino acids that could be operationally distinguished by aaRS enzymes. To account for this, a parallel alignment of the amino acid sequences of the highly conserved structural cores of Class I and II enzymes \cite{o2003evolution,Popinga}, representative of the primordial structure from which all such enzymes are derived, has been assembled and then analysed using substitution matrices of different sizes for different evolutionary epochs (see Section 4.2). While there is still considerable uncertainty about the exact structure of this phylogeny, the tree we use here is a good candidate for explaining the empirically discernible history of aaRS divergence.  It is notable that except for the position of the histidyl-tRNA synthetase in the pseq tree, its clade structure conforms to the observed sub-classification of aaRS enzymes according to their structure and biochemistry \cite{o2003evolution}.

For subsequent analysis we constructed trees in which the split into Classes I and II was followed by a further binary split, within each Class, between aaRS enzymes with specificities for polar amino acids \{arg, asn, asp, glu, gln, his, lys, ser, thr, tyr\} and nonpolar amino acids \{ala, cys, gly, ile, leu, met, phe, pro, trp, val\}.  These trees were labelled ``I:II-pol''.  Note that I:II-pol trees have a ``ten-(six-four)'' structure because, by our classification, four Class I amino acids \{arg, glu, gln, tyr\} are polar and four Class II amino acids \{ala, gly, phe, pro\} are nonpolar.  We also considered completely hypothetical ten-(six-four) trees in which there was an initial split on the basis of amino acid polarity and then a further binary split, not only within the set of aaRS with for specificity polar amino acids, but also within the nonpolar grouping, according to the actual biological Class of aaRS.  These trees were labelled ``pol-I:II''. In Figs. \ref{histPolAars} and \ref{histAarsPol}, results for these two types of trees are compared with the tree types built with fewer constraints (random, ten-ten, I:II).   

\begin{figure}
\centering
\includegraphics[scale = 0.6]{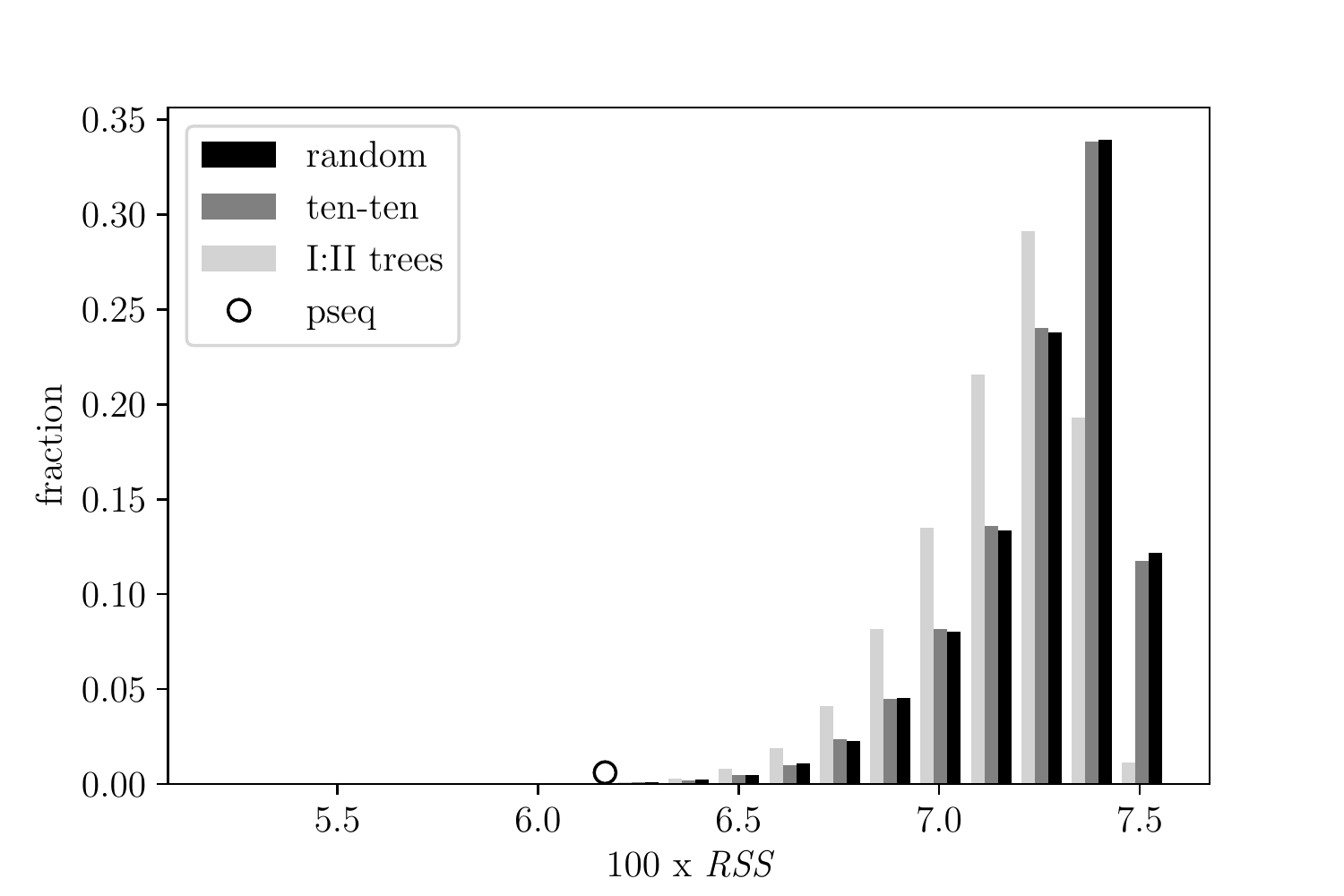} 
\caption{Histograms of $RSS$ values for comparison of an empirical aaSM (LG) with various distributions of parameterized aaSMs derived from specified types of trees representing hypothetical aaRS phylogenies.  The open circle shows the comparison with an aaRS tree obtained through phylogenetic analysis of actual aaRS sequences. The ``pseq'' point shows the result for a tree deduced from phylogenetic fitting to aaRS protein sequence data: (((M,(I,(V,L))),((Y,W),(((E,Q),C),R))),(((F,H),(K,(D,N))), (((S,P),(T,G)),A))); in Newick format and aaRS types labelled using the one letter designation for amino acids.}
\label{hist1}
\end{figure}

\subsection{Parameterized fitting of tree-derived matrices to LG aaSM}

The first result evident from Fig. \ref{hist1} is that imposing the ``ten-ten'' constraint on the random selection of bifurcating branches has virtually no effect on the overall distribution of distances \textit{RSS} that the parameterized matrices (black and dark grey histograms in Fig. 2) have from the empirical matrix (LG). The result of a t-test  showed that there was no significant difference between the mean of the two distributions ($ t = 0.843$, $ p=0.399$). However, when the ten-ten split is constrained to correspond to the observed ``I:II'' separation of amino acid specificities dictated by the canonical Class I and II aaRS types, then there is a marked improvement in the extent to which the empirical matrix is represented (light grey histogram). The t-test results show a significant difference between the means of the distribution for I:II trees compared with both random trees ($ t = 34.4 $, $ p <0.001$) and ten-ten trees ($ t=35.5 $, $p <0.001 $).  However, it is striking that the pseq tree, recently derived by joint phylogenetic analysis of actual Class I and II aaRS sequences \cite{Popinga}, is an outlier in the distribution of I:II trees to which it belongs.   That this empirically derived aaRS phylogeny is an outlier in the distribution of I:II trees, fitting the empirical aaSM far better than most others, indicates that aaRS evolution has catered to far more refined aspects of amino acid chemistry than those that could be differentiated by the separation of aaRS specificities according to the primordial I:II Class separation.

\begin{figure} 
\centering
\includegraphics[scale = 0.6]{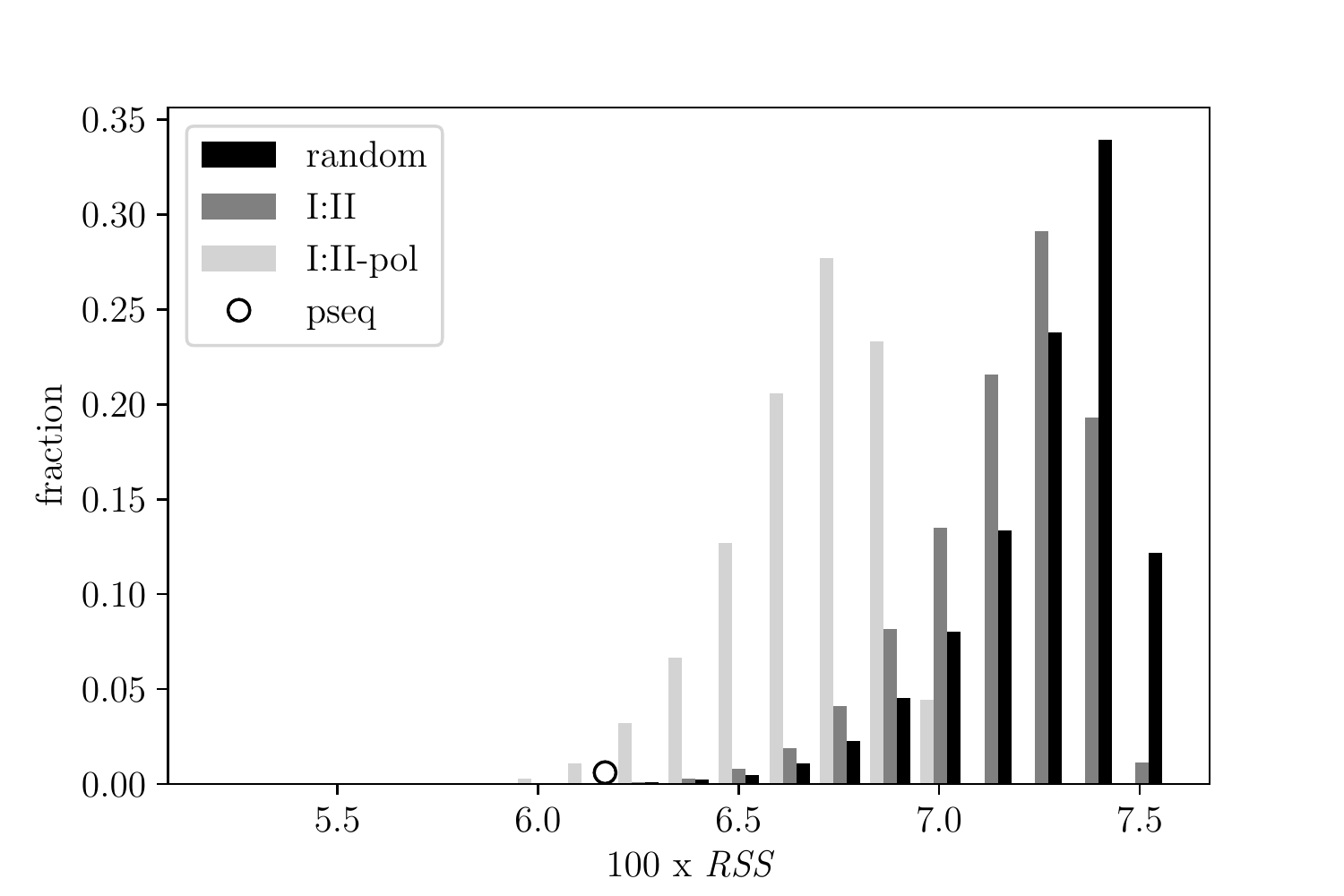}
\caption{Histograms of $RSS$ values for comparison of an empirical aaSM (LG) with distributions of parameterized aaSMs derived from Class-split (I:II) aaRS phylogenies and the subset of these trees (I:II-pol) in which the aaRS specificities of enzymes in each class was further split into subsets of higher and lower polarity. The branching pattern of the “pseq” aaRS phylogeny is defined in the caption of Figure \ref{hist1}.}
\label{histPolAars}
\end{figure}

The major effect of polarity on the specificity of aaRS selectivity can be seen in Fig. \ref{histPolAars}.  The distribution of \textit{RSS} values for I:II-pol trees shows that phylogenies, in which aaRS sequence similarity within each Class depends primarily on the polarity of the amino acid recognised, correspond to parameterized aaSM matrices that fit the experimental LG matrix much better than the bulk of I:II trees. The t-test that compared the mean \textit{RSS} for these two types of tree gave the results $ t = 166 $ and $ p <0.001 $. 
We found that the parameterized aaSM matrices corresponding to I:II-pol trees provided a statistically significant improvement over the parameterized aaSM matrices corresponding to I:II trees ($F = 10.45$, $p = 0.0024$).
The p-value was calculated by finding the reverse percentile from a simulation-derived null distribution with $N=100,000$ entries (see Methods section for details).
  However, it is notable that the performance of the tree (pseq) that resulted from bioinformatic analysis of amino acid sequences of aaRS proteins is still significantly superior to the bulk of I:II trees, even those in the I:II-pol sub-distribution.

\begin{figure}
\centering
\includegraphics[scale = 0.6]{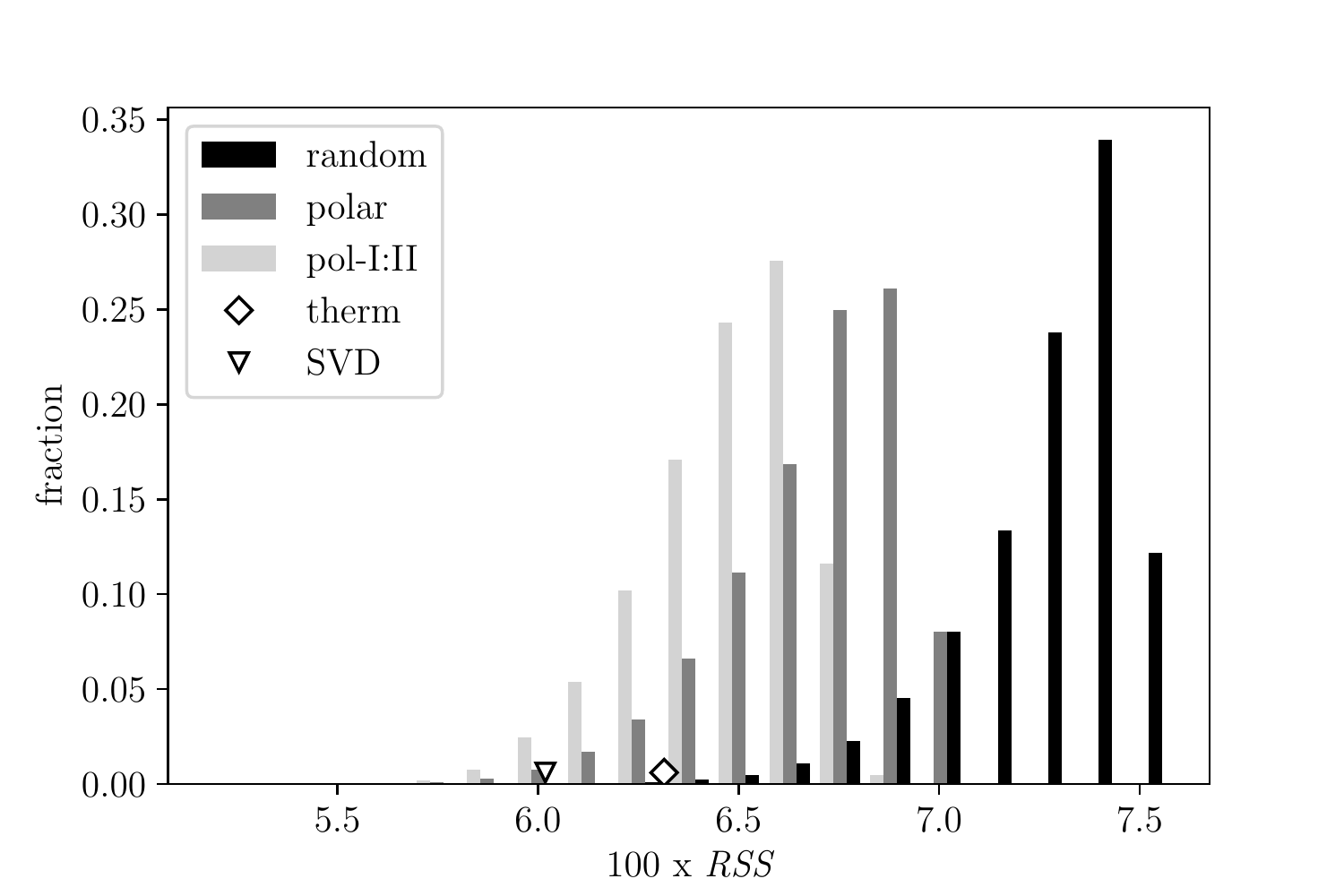}

\caption{Histograms of $RSS$ values for comparison of an empirical aaSM (LG) with distributions of parameterized aaSMs derived from aaRS phylogenies initially ``ten-ten'' split according to amino acid polarities (polar) and the subset of these trees in which the aaRS specificities of enzymes in each ``polar'' division were further split into subsets according to their canonical aaRS I:II Class division.  The ``therm'' point shows the result obtained from thermodynamic fitting according to the forms of Eqs. (2) and (3); and the ``SVD'' point signifies fitting with three vectors according to Eq. (4).}
\label{histAarsPol}
\end{figure}

In order to investigate the putative relationship between aaRS phylogeny and the aaSM derived from protein structural homology  more deeply, we constructed completely hypothetical pol-I:II trees based on an initial bifurcation into enzymes with specificities for polar and nonpolar amino acids  (``polar'' type trees) followed by a further bifurcation, within each branch, based on the canonical I:II split observed across the tree of life.  
We found that the additional two parameters in the parameterized aaSM matrices corresponding to pol-I:II trees provided a statistically significant improvement over the parameterized aaSM matrices corresponding to pol trees ($F = 5.71$, $p = 0.0343$).
The p-value was calculated by finding the reverse percentile from a simulation derived null distribution with $N=100,000$ entries (see Methods section for details).
This result demonstrates that even after the dominant effect of polarity is taken into account, aaRS class still has relevant information to ``add'', improving the fit of the trees to the empirical aaSM.


\subsection{Thermodynamic modelling}
We further examined the dominant role of amino acid polarity in determining the form of the empirical aaSM by investigating how well the matrix could be represented in terms of proxy measures of the free energy change $\Delta G_{ij}$ associated with the substitution of one amino acid $i$ for another $j$, or vice versa, in the structure of folded proteins.  
Previously, Wolfenden and Carter \cite{WolfendenCarter2015,carter2015trna} showed how these free energy differences could be variously broken down into terms describing more elementary transfers of a single amino acid sidechain $i$ from the vapour phase ($v$) into cyclohexane ($c$) or water ($w$), $\Delta G_{i}$($v>c$) and $\Delta G_{i}$($v>w$), respectively, and between the two solvents, $\Delta G_{i}$($w>c$)  =  $ \Delta G_{i}$($v>c$) $-$ $\Delta G_{i}$($v>w$).  
We therefore defined a $20 \times 20$ matrix $\+\Delta_{\textbf{\emph{vc}}}(ij)=|\Delta G_i(v$$>$$c)-\Delta G_j(v$$>$$c)|$ corresponding to the unsigned magnitude of the free energy needed to replace amino acid $i$ with $j$, taking $i$ from the cyclohexane solvent into the vapour phase and replacing it with amino acid $j$ taken from the vapour phase. The corresponding matrix $\+\Delta_{\textbf{\emph{vw}}}(ij)$ was for the same measure but with water instead of cyclohexane as the solvent.  A third matrix $\+\Delta_{\textbf{\emph{wc}}}(ij)=|\Delta G_i(w$$>$$c)-\Delta G_j(w$$>$$c)|$ corresponds to the  difference, between amino acids $i$ and $j$, of transferring the amino acid from water to cyclohexane.  Thus, the $\+\Delta$ matrices represent differences between the characteristic partitioning of individual amino acids across vapour-solvent and solvent-solvent boundaries.  Our approach is similar to that of {\v{S}}tambuk \emph{et al}. \cite{vstambuk2016miyazawa}. 

Considering terms involving all possible monadic and dyadic matrix forms, it was found that the equation that best fitted the empirical LG aaSM was  

\begin{align}
\begin{split}
\+{LG_{calc}} = & 2.63 - 0.553 \+\Delta_{\textbf{\emph{vc}}} - 0.248 \+\Delta_{\textbf{\emph{wc}}} \\ & + 0.0628 \+\Delta_{\textbf{\emph{vc}}} \+\Delta_{\textbf{\emph{wc}}} \label{deltaFit}
\end{split}
\end{align}

\noindent where the matrix product $\+\Delta_{\textbf{\emph{vc}}} \+\Delta_{\textbf{\emph{wc}}}$ was determined using term-by-term multiplication for each $ij$ entry. The calculated matrix fitted the empirically derived LG matrix with an \textit{RSS} value of 0.0666, a correlation coefficient of $ r = 0.485 $ and coefficient of determination $ r^2 = 0.235 $.  When the volume $V$ of the amino acid sidechains, taken from \cite{richards1977areas}, was allowed as a variable, the equation that best fitted the data was found to be

\begin{align}
\begin{split}
\+{LG_{calc}} = & 3.28 - 0.304 \+\Delta_{\textbf{\emph{vw}}} - 0.0310 \+\Delta_{\textbf{\emph{V}}} \\ &  + 0.00320 \+\Delta_{\textbf{\emph{vw}}} \+\Delta_{\textbf{\emph{V}}}  \label{deltaVolFit}
\end{split}
\end{align}

\noindent where $\+\Delta_{\textbf{\emph{V}}}(ij) = |V_i - V_j| $. The goodness of fit of these two representations of the LG matrix in terms of thermodynamic parameters (Eqs. \ref{deltaFit} and \ref{deltaVolFit}) was practically identical and is compared with that of the pol-I:II distribution in Fig. \ref{histAarsPol}, denoted there by ``therm''.

\subsection{Singular value decomposition}

In order to provide comparison of our results with the extensive work of the Pokarowski group \cite{pokarowski2007ideal} we investigated one purely parametric representation of the LG aaSM, but instead of using the protocol of that group we followed the approach of Zimmerman \cite{Zimmerman} and relied on singular value decomposition (SVD) of the empirical LG matrix.  SVD analysis consists  of finding a set of 20-dimensional vectors ${\textbf{\emph{x}}, \textbf{\emph{y}}, \textbf{\emph{z}}} ...$, each representing an array of values of an arbitrary parameter for each amino acid $i=1$$:$$20$, that best fit the target matrix according to the relationship 

\begin{equation}
\+{LG_{calc}}(ij) = x_ix_j + y_iy_j + z_iz_j +\ldots .
\end{equation}

\noindent As the number of SVD vectors increased from 1 to 3, the \textit{RSS} value of the fit achieved decreased from 0.0743 to 0.0659 to 0.0602.  The final value is shown in Figure \ref{histAarsPol} as representative of this purely statistical approach to the numerical representation of the aaSM.

\section{Discussion}
\label{sec:2}
The elements of the aaSM provide an inverse ``mean field'' measure of the selection pressure against nonsynonymous mutations in coding sequences $-$ high values of exchangeability indicate changes in amino acids that functional proteins tolerate easily; low exchangeability values indicate a lack of general tolerance.  Thus, the aaSM is a key to our understanding of evolution of phenotypes at the molecular level.  In this study we have extended this insight to enquire into the earliest stages of the ``evolution of evolvability'' \cite{Wagner1,Wagner2} by looking for traces of the pathway along which the refined genotype-to-phenotype mapping of the standard genetic code first developed.  We have found that the structure of the aaSM is indeed consistent with an understanding of early molecular evolution as a self-organising process: proteins and nucleic acids, especially those involved in protein synthesis, spontaneously built up progressively more refined maps of one another's properties, starting from virtually nothing.

A potential criticism of our approach is that its reasoning is circular: the observed aaSM is explained in terms of parameters derived from an aaRS phylogeny that reflects the progressive differentiation of amino acids according to their chemical similarities and differences, the very property that entries in the aaSM necessarily measure, given that proteins are under selective pressure to maintain vital functions.  However, such criticism fails to take into account the array of evidence from disparate sources that independently corroborates the narrative through which we can now bind the chemical properties of amino acids to the evolution of the genetic code.  It also fails to take into account the intrinsic circularity in nature's logic: similarities and differences between amino acid sidechains in sequence positions along the peptide backbones of the Class I and II aaRS structural cores are used to define what similarities and differences are operationally useful to confer catalytic specificity on aaRS enzymes, especially the differentiation of amino acids according to their various chemical properties. With this in mind we can now make sense of our results within the context of the evolutionary processes that allowed a functional map of the chemical properties of amino acids to become embedded as information in the table of nucleotide triplet codons.

We began with the observation that, for coding purposes, the means of identifying amino acids is accomplished by aaRS enzymes that are divided into two protein superfamilies, which bear no obvious structural similarity to one another and do not have a common origin, save the possibility that their progenitors were encoded on the complementary strands of a single nucleic acid gene \cite{RodinOhno1995}. Nevertheless, each superfamily Class has 10 members of varying amino acid specificity, and that specificity, by whatever measure, appears to have been distributed between Class I and II aaRS enzymes in a very haphazard manner.  The first question we asked was whether the evenness in the division of amino acid specificities between the Class I and II aaRS enzymes was a significant factor bearing on the ability of coding to distinguish amino acids according to their functional chemical properties and thus provide a basis for the reliable production of proteins with high functional specificity.  

Unsurprisingly, parameterized aaSMs derived from ``ten-ten'' aaRS phylogenies, in which the 20 enzymatic specificities were divided into equal classes, provided a distribution of fits to the empirical aaSM that was barely distinguishable from the distribution derived from all possible phylogenies.  The ten-ten tree-shape constraint has little effect.  However, when the choice of amino acids represented by the choice of aaRS specificities on either side of the ten-ten split was made exactly according to the split between the naturally occurring Class I and II enzymes (I:II trees), the parameterized fit to the empirical aaSM was significantly improved.  Clearly, the historically established division of aaRS specificities into Classes I and II aaRS reflects, at least to some degree, how the chemical properties of amino acids allow them to substitute for one another in the folded structure of functional proteins.  

The main difference between the amino acids recognised by the two aaRS Classes is their average size \cite{carter2015trna}, but the I:II split is a relatively minor factor in the overall interaction processes supporting coding through the recognition of amino acids by aaRSs. The pseq tree, obtained from aaRS sequence data, provides an extraordinarily good parameterized representation of the empirical aaSM compared with the large bulk of possible trees of the I:II type.  This indicates how well more detailed features of aaRS evolution, not just the separation of the enzymes into Classes I and II, have produced similar enzymes for recognising amino acids with similar properties according to the criterion of their substitutability within folded proteins.  It also gives us reason to believe that the more detailed clade structure of aaRS enzymes, beyond their canonical division into Classes I and II, reflects the detailed evolutionary process whereby the genetic code's map of the chemical properties of amino acids onto the codons' six bits of information developed step by step.    

As is abundantly clear from consideration of the forces between amino acid sidechains that maintain the secondary and tertiary structures of functional proteins, polarity (or hydrophobicity) is the most general property determining the role and substitutability of individual amino acids in folded proteins.  This was born out fully (Fig. \ref{histPolAars}) in the analysis of I:II-pol trees, which were first split according to aaRS Class (I:II) and then subjected to a second split, within each Class, so that polar and non-polar amino acids were consigned to separate branches.  

The overall improvement of the fit of parameterized aaSM derived from I:II-pol trees, compared with that obtained from simple I:II trees, strongly corroborates the scenario of code development proposed in \cite{carter2015trna,carter2017interdependence}: the primitive operational code matched tRNA acceptor-stem identity elements to amino acids of broadly different sizes, but the advent of the tRNA anticodon loop \cite{DiGiulio1995,DiGiulio2004} allowed each size-class to be differentiated much more finely, now according to the orthogonal chemical property of electric polarity.  The emergent system for biochemical recognition of base-triplet codons using the 4-letter {A,C,G,T} nucleotide alphabet provided a discrete information cube with sides of length 2 bits within which the chemical properties of amino acids could be mapped in a way that optimized robust protein functionality. 

Because the aaRS catalysts which effect the mapping between codons and amino acids are themselves proteins, we must view the evolution of the code as a self-optimising process.  The aaRS sequence-derived pseq tree \cite{Popinga} gives a parameterized fit to the empirical aaSM that is vastly superior to that of most other I:II-pol trees. This demonstrates that distinctions between amino acids very much finer than those achievable simply by their placement on the scale of polarity (or, inversely, hydrophobicity) went into the evolutionary process whereby aaRS enzymes learned how to use arrangements of amino acid sidechains to create binding sites specific for the recognition of individual sidechains.

We must emphasise that the results of our investigations of aaSM form can only corroborate, not provide a means of deducing or proving, any particular theory or narrative concerning code evolution.  On the other hand, the corroboration is strong.  The idea that a biochemically ``operational code'' preceded the full genetic code in which nucleotide triplets serve as 6-bit tokens of information is born out by the small but significant proportion of pol-I:II trees (Figure \ref{histAarsPol}) that give a better parameterized fit to the empirical aaSM than the sequence-derived aaRS phylogenetic tree  (pseq).  Apparently, if the code had started out with the amino acids neatly sorted according to their functionally dominant property of polarity, then aaRS differentiation and code expansion would have provided a path that was more straightforward and more easily extended in the direction of the fit-for-purpose aaSM, which we may call ``modern'' even though it is presumably ~3.5 billion years old.

The fact that these pol-I:II trees are fictional, whereas the real pseq tree is of the I:II-pol type, reflects a ``frozen accident'' aspect of the manner in which the stepwise process of aaRS phylogeny refined the codon representation of amino acid sidechain chemistry.  The initial operational code was suboptimal, mostly using crude groove recognition to distinguish features of the three uppermost base pairs of the primitive tRNA acceptor stem \cite{carter2018hierarchical}.  In the event, this primitive mapping of amino acid properties using acceptor stem ``identity elements'' could not be entirely overwritten when the anticodon loop became available as a target of selection for a system of coding based on discretely recognisable nucleotide triplets.  The same conclusion has been arrived at using a quite different definition of code optimization \cite{Facchiano2018}.  The final polarity-based optimization of coding, instantiated in the modern aaSM and employing bits of information available in the anticodon loop, could only be achieved by circumventing rather than completely overwriting the prior effects of entrenched groove-recognition patterns pairing Class I and II aaRS enzymes with different tRNA acceptor stems.

In the end, entrenchment of the poorly differentiated representation of amino acid chemistry that the operational code afforded did not hinder the emergence of a highly refined code.  The reflexive definition of the chemical properties of amino acids through their own functional utilization in aaRS enzymes drove the evolution of coding close to the limit of resolution for systems that employ proteins as machines that recognise their own building blocks \cite{carter2017interdependence}.  The map of amino acid similarity which evolution created reflects its own utility  \cite{StephensonFreeland}.




\section{Methods}
\label{sec:3}

\subsection{Matrix normalization}
In order to facilitate easy comparison between different ways of approximating aaSM matrices, we employed a normalised representation of the chosen matrix.  Normalisation was achieved by ensuring that the off-diagonal entries of the symmetric exchangeability matrix summed to unity.  

\subsection{Protein sequence (pseq) tree}

The pseq tree (Figs. 3 and 4) is a representative, candidate tree of high posterior probability obtained as a preliminary result in the study of Popinga et al. \cite{Popinga}.  That phylogenetic analysis is based on a sequence alignment of amino acids found in conserved “scaffold” positions of the aaRS structures of each Class.  The alpha carbon atoms of the approximately 100 amino acids in these positions occupy structurally homologous locations across each entire Class of aaRS from all organisms in the tree of life.  The phylogenetic analysis of the scaffold position alignment is conducted using a purpose-built, online-available package \cite{Popinga} built onto the BEAST2 software \cite{bouckaert2019beast}.  Consonant with the description in Section 2.1 the package evaluates joint trees of Class I and II aaRS for their goodness-of-fit to aligned amino acid sequence data by constructing and making use of amino acid substitution matrices of reduced dimension for different epochs according to the number of coalescence steps passed through in the ascent from the tips to the root of any candidate tree. A more detailed description and justification of the methodology is in preparation, and the acquisition and analysis of more certain and complete results is in progress.   

\subsection{The F-test}

When a set of $n$ data points is represented by two different models, and model 1 with $p_{1}$ parameters is nested within model 2 with $ p_{2} > p_{1} $ parameters, the F-test relies on the distribution of the statistic

\begin{equation} F = \scalebox{1.5}{${\tfrac{RSS_{1}-RSS_{2}}{p_{2}-p_{1}}}/{\tfrac{RSS_{2}}{n-p_{2}-1}}$} \label{Fformula} \end{equation}

\noindent to assess the statistical legitimacy of preferring the more specialised second model.  For each model the residual sum of squares is as defined in Eq. (1). Note that, the best fitting value of each parameter is found by simply averaging all the values in equivalent positions in the empirically derived matrix. In our case we consider a specialised parametrized model aaSM relative to a broader parametrized model in which it is nested, for instance the parametrized models defined by I:II-pol trees nested within the broader group of parameterized models defined by I:II trees.

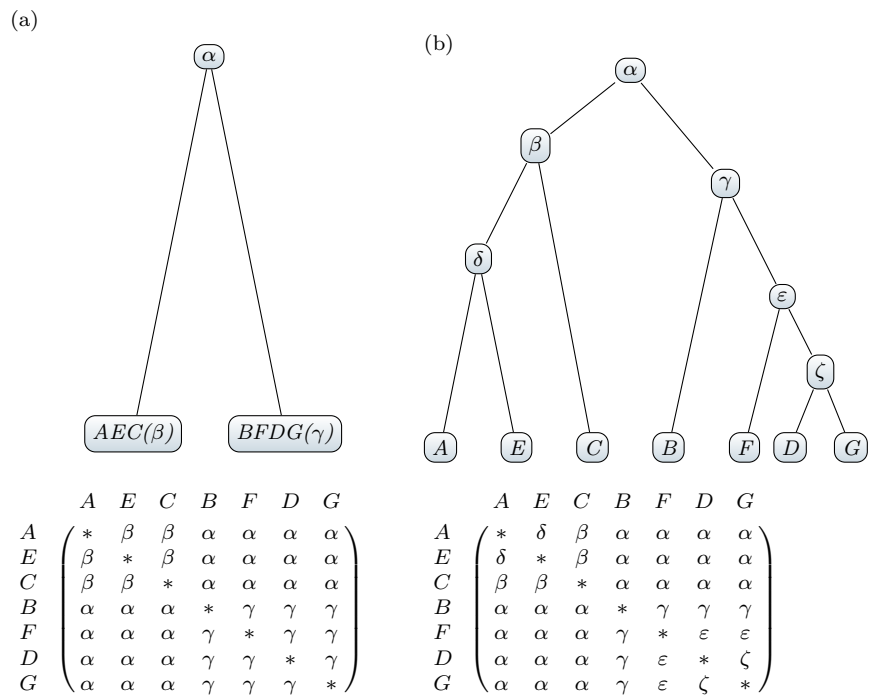
\begin{figure}

(a) \\ \begin{minipage}{0.45\textwidth}
\centering
\begin{tikzpicture}[level 1/.style={sibling distance=20mm},level 2/.style={sibling distance=50mm},
  every node/.style = {shape=rectangle, rounded corners,
    draw, align=center,
    top color=white, bottom color=indigo(dye)!20}]]
  \node {$ \alpha $}
    child[level distance = 5cm] { node {\em AEC($ \beta $)} }
    child[level distance = 5cm]{ node {\em BFDG($ \gamma $)}};
\end{tikzpicture}

\end{minipage}
\begin{minipage}{0.45\textwidth}

(b) \\ \begin{tikzpicture}[level 1/.style={sibling distance=25mm},level 2/.style={sibling distance=15mm},level 3/.style={sibling distance=10mm},level 4/.style={sibling distance=8mm},
  every node/.style = {shape=rectangle, rounded corners,
    draw, align=center,
    top color=white, bottom color=indigo(dye)!20}]]
  \node {$ \alpha $}
    child[level distance = 1cm] { node {$ \beta $}
    	child[level distance = 1.5cm] { node {$ \delta $}
        	child[level distance = 2.5cm] {node {\em A} } 
            child[level distance = 2.5cm] {node {\em E} } }
        child[level distance = 4cm] { node {\em C} 
        	 } }
    child[level distance = 1.5cm]{ node {$ \gamma $} 
    	child[level distance = 3.5cm]{ node {\em B} }
        child[level distance = 1.5cm]{ node {$ \varepsilon $} 
        	child[level distance = 2cm]{node{\em F}} 
        	child[level distance = 1cm]{ node {$ \zeta $} 
            	child[level distance = 1cm] {node{\em D}}
                child[level distance = 1cm] {node{\em G}}  } 
        	}
          };
\end{tikzpicture} 

\end{minipage}

\begin{minipage}{0.45\textwidth}
\[ \bordermatrix{ & A & E & C & B & F & D & G \cr
      A \phantom{i} & * & \beta & \beta & \alpha & \alpha & \alpha & \alpha  \cr
      E \phantom{i} & \beta & * & \beta & \alpha & \alpha & \alpha & \alpha  \cr
      C \phantom{i} & \beta & \beta & * & \alpha & \alpha & \alpha &  \alpha \cr
      B \phantom{i} & \alpha & \alpha &  \alpha & * & \gamma & \gamma & \gamma  \cr
      F \phantom{i} & \alpha & \alpha & \alpha & \gamma & * & \gamma & \gamma \cr 
      D \phantom{i} & \alpha & \alpha & \alpha & \gamma & \gamma & * & \gamma \cr 
      G \phantom{i} & \alpha & \alpha & \alpha & \gamma & \gamma & \gamma & *
      } \qquad
\]
\end{minipage}
\begin{minipage}{0.45\textwidth}
\[  \bordermatrix{ & A & E & C & B & F & D & G \cr
      A \phantom{i} & * & \delta & \beta & \alpha & \alpha & \alpha & \alpha  \cr
      E \phantom{i} & \delta & * & \beta & \alpha & \alpha & \alpha & \alpha  \cr
      C \phantom{i} & \beta & \beta & * & \alpha & \alpha & \alpha &  \alpha \cr
      B \phantom{i} & \alpha & \alpha &  \alpha & * & \gamma & \gamma & \gamma  \cr
      F \phantom{i} & \alpha & \alpha & \alpha & \gamma & * & \varepsilon & \varepsilon \cr 
      D \phantom{i} & \alpha & \alpha & \alpha & \gamma & \varepsilon & * & \zeta \cr 
      G \phantom{i} & \alpha & \alpha & \alpha & \gamma & \varepsilon & \zeta & *
      } \qquad
\]
\end{minipage}

\caption{An seven-letter example of two nested models. The tree on the right is a refinement of the tree on the left. The parameterized model defined by the tree on the left, which has two parameters, is nested within the parametrized model on the right, which has 6 parameters. The model on the right will necessarily provide a better fit to any given empirically derived matrix than the model on the left.}
\label{FtestTrees}
\end{figure}

Two trees are formally defined to be nested if their rate matrices become equal when additional constraints are imposed on the one with the greater number of free parameters.  When the parameterised aaSMs corresponding to two nested trees are both fitted to the same empirically derived aaSM, the one which is more refined and therefore has the greater number of free parameters, will fit the empirical matrix at least as well as the less refined tree. Note that in a tree which is not fully resolved, amino acids that are grouped together on the same leaf are assumed to have the same rate of exchange with any other amino acid not on their common leaf. This being the case, the less resolved tree can still be fitted to the same aaSM as a fully resolved tree. An example is shown in Figure \ref{FtestTrees}, which shows two nested trees with $n = 7$ and their corresponding rate matrices (cf Figure 1b). In this example, the constraints  $ \beta = \delta $ and $ \gamma = \varepsilon = \zeta $ equate the two matrices.

\begin{figure}[h]

\begin{minipage}{0.32\textwidth} (a)
\begin{tikzpicture}[
  every node/.style = {shape=rectangle, rounded corners,
    draw, align=center,
    top color=white, bottom color=indigo(dye)!20}]]
  \node {}
    child[level distance = 1.5cm, sibling distance = 20mm] { node {} 
    	child[level distance = 1.5cm, sibling distance = 1.8mm]{}
        child[level distance = 1.5cm, sibling distance = 1.8mm]{}
        child[level distance = 1.5cm, sibling distance = 1.8mm]{}
        child[level distance = 1.5cm, sibling distance =1.8mm]{}
        child[level distance = 1.5cm, sibling distance = 1.8mm]{}
        child[level distance = 1.5cm, sibling distance = 1.8mm]{}
        child[level distance = 1.5cm, sibling distance = 1.8mm]{}
        child[level distance = 1.5cm, sibling distance = 1.8mm]{}
        child[level distance = 1.5cm, sibling distance = 1.8mm]{}
        child[level distance = 1.5cm, sibling distance = 1.8mm]{} }
    child[level distance = 1.5cm, sibling distance = 20mm] { node {} 
    	child[level distance = 1.5cm, sibling distance = 1.8mm]{}
        child[level distance = 1.5cm, sibling distance = 1.8mm]{}
        child[level distance = 1.5cm, sibling distance = 1.8mm]{}
        child[level distance = 1.5cm, sibling distance = 1.8mm]{}
        child[level distance = 1.5cm, sibling distance = 1.8mm]{}
        child[level distance = 1.5cm, sibling distance = 1.8mm]{}
        child[level distance = 1.5cm, sibling distance = 1.8mm]{}
        child[level distance = 1.5cm, sibling distance = 1.8mm]{}
        child[level distance = 1.5cm, sibling distance = 1.8mm]{}
        child[level distance = 1.5cm, sibling distance = 1.8mm]{} 
    }  ;
\end{tikzpicture}
\end{minipage}
\begin{minipage}{0.32\textwidth} (b)
\begin{tikzpicture}[
  every node/.style = {shape=rectangle, rounded corners,
    draw, align=center,
    top color=white, bottom color=indigo(dye)!20}]]
  \node {}
    child[level distance = 1cm, sibling distance = 2cm] { node {}
      child[sibling distance = 9mm]{node {} 
      	child[sibling distance = 1.8mm]{}
        child[sibling distance = 1.8mm]{}
        child[sibling distance = 1.8mm]{}
        child[sibling distance = 1.8mm]{}
        child[sibling distance = 1.8mm]{}
        child[sibling distance = 1.8mm]{}
      }
      child[sibling distance = 9mm]{node {}
      	child[sibling distance = 1.8mm]{}
        child[sibling distance = 1.8mm]{}
        child[sibling distance = 1.8mm]{}
        child[sibling distance = 1.8mm]{}
      }}
    child[level distance = 1cm, sibling distance = 1.6cm] {node {}
      child[sibling distance = 9mm]{node {} 
      	child[sibling distance = 1.8mm]{}
        child[sibling distance = 1.8mm]{}
        child[sibling distance = 1.8mm]{}
        child[sibling distance = 1.8mm]{}
      }
      child[sibling distance = 9mm]{node {}
      	child[sibling distance = 1.8mm]{}
        child[sibling distance = 1.8mm]{}
        child[sibling distance = 1.8mm]{}
        child[sibling distance = 1.8mm]{}
        child[sibling distance = 1.8mm]{}
        child[sibling distance = 1.8mm]{}
      }}  ;
\end{tikzpicture}
\end{minipage}
\caption{The structure of the randomly selected nested pairs of trees used to generate the null distribution shown in Fig. \ref{Fdistribution}. \label{treesForFtest}}
\end{figure}
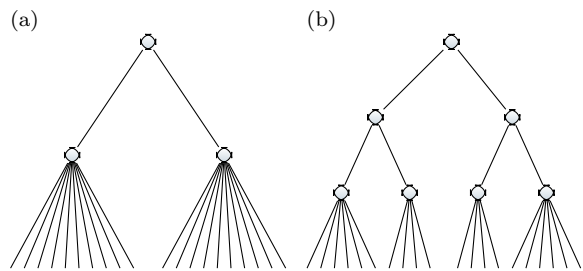

\begin{figure}
\includegraphics[scale = 0.6]{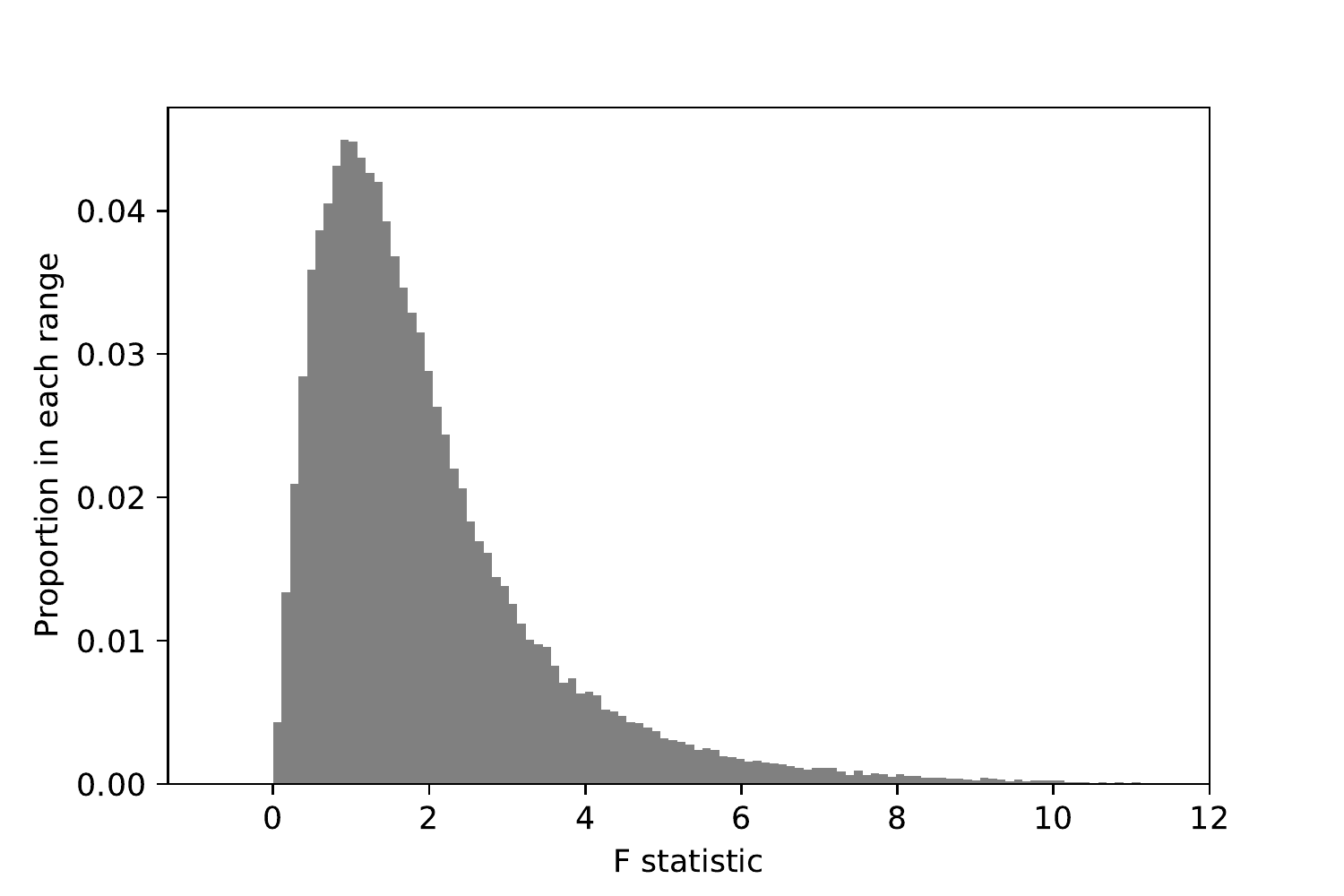}
\caption{Distribution of the F-statistic for $N=1,000,000$ randomly selected nested pairs of trees with the structure shown in Figure \ref{treesForFtest}. Treating this as a null distribution the critical value of the F-statistic ($\alpha = 0.05$) is 5.094.}
\label{Fdistribution}
\end{figure}

We wish to compare the value of the F statistic to a null distribution generated under the hypothesis that the data came from the simpler model. The F-distribution rests on the assumption that the response data points are independent of each other, which is not the case here due to the matrix structure of the data. This means that we cannot use the F-distribution as our null distribution. The simplest way forward is to generate null distributions by simulation so as to avoid having to determine the statistically relevant degrees of freedom. The null distribution of interest is one which compares two nested trees with 20 leaves. 
For each entry in the null distribution we randomly generated a tree with the ten-ten structure (Figure \ref{treesForFtest} (a)). Then for each set of ten labels we randomly subdivide them into groups of size four and six giving a tree with the structure shown in Figure \ref{treesForFtest} (b). The parameterized models associated with these two trees are nested. Equation \ref{Fformula} can then be used to calculate an F statistic for the pair of models.
To estimate the null distribution we generate $N=100,000$  pairs of nested trees with the structure shown in Figure \ref{treesForFtest}.  The distribution of the F scores is shown in Figure \ref{Fdistribution}. This distribution conforms to the expected shape of an F-distribution. The 95th percentile of scores was recorded to be $ 5.094 $ which, for a type-I error probability of $ \alpha = 0.05 $, is the F critical value for this distribution.

\subsection{Thermodynamic analysis}
From \cite{carter2015trna} we obtained values of three related thermodynamic parameters characterising the phase equilibria of amino acid sidechains between different chemical environments: vapour to water transfer equilibrium ($v$$>$$w$) representing hydrophilic character, vapour to cyclohexane transfer equilibrium ($v$$>$$c$) representing hydrophobic character, and water to cyclohexane transfer equilibrium ($w$$>$$c$) representing polarity. The pairwise differences between these parameters for amino acids $i$ and $j$ were used to generate the $ (20 \times 20) $ matrices  $\+\Delta_{\textbf{\emph{vw}}}(ij)$, $\+\Delta_{\textbf{\emph{vc}}}(ij)$ and $\+\Delta_{\textbf{\emph{wc}}}(ij)$, which were used as predictors for aaSMs (Eq 2). Dyadic terms in the regression equation were evaluated using term-by-term matrix multiplication.
Mantel regression was used through the R ecodist package \cite{Rprogram,ecodist}.

%
%

\begin{acknowledgements}
PRW thanks the Alexander von Humboldt Foundation for its continual support; both Peter Stadler and Alexei Drummond for their encouragement, more than a decade ago, to carry out this study; Andrew Torda for advice about substituion matrices; and Charlie Carter for constant helpful correspondence and discussions.

\end{acknowledgements}

%
%

\bibliographystyle{spmpsci}      
\bibliography{manuscript.bib}   

\begin{thebibliography}{10}
\providecommand{\url}[1]{{#1}}
\providecommand{\urlprefix}{URL }
\expandafter\ifx\csname urlstyle\endcsname\relax
  \providecommand{\doi}[1]{DOI~\discretionary{}{}{}#1}\else
  \providecommand{\doi}{DOI~\discretionary{}{}{}\begingroup
  \urlstyle{rm}\Url}\fi

\bibitem{Atchley2005}
Atchley, W.R., Zhao, J., Fernandes, A.D., Dr{\"u}ke, T.: Solving the protein
  sequence metric problem.
\newblock Proc. Natl. Acad. Sci. USA \textbf{102}, 6395--6400 (2005)

\bibitem{Bashford1998}
Bashford, J., Tsohantjis, I., Jarvis, P.: A supersymmetric model for the
  evolution of the genetic code.
\newblock Proc. Natl. Acad. Sci. USA \textbf{95}, 987--992 (1998)

\bibitem{bernhardt2008evidence}
Bernhardt, H.S., Tate, W.P.: Evidence from glycine transfer {RNA} of a frozen
  accident at the dawn of the genetic code.
\newblock Biology Direct \textbf{3}(1), 53 (2008)

\bibitem{bouckaert2019beast}
Bouckaert, R., Vaughan, T.G., Barido-Sottani, J., Duch{\^e}ne, S., Fourment,
  M., Gavryushkina, A., Heled, J., Jones, G., K{\"u}hnert, D., De~Maio, N.,
  et~al.: {BEAST} 2.5: An advanced software platform for {B}ayesian
  evolutionary analysis.
\newblock PLoS computational biology \textbf{15}(4), e1006650 (2019)

\bibitem{caetano2013structural}
Caetano-Anoll{\'e}s, G., Wang, M., Caetano-Anoll{\'e}s, D.: Structural
  phylogenomics retrodicts the origin of the genetic code and uncovers the
  evolutionary impact of protein flexibility.
\newblock PLoS One \textbf{8}(8), e72225 (2013)

\bibitem{CarterLife}
Carter, C.W.: What {RNA} world? why a peptide/{RNA} partnership merits renewed
  experimental attention.
\newblock Life \textbf{5}, 294--320 (2015)

\bibitem{carter2017interdependence}
Carter, C.W., Wills, P.R.: Interdependence, reflexivity, fidelity, impedance
  matching, and the evolution of genetic coding.
\newblock Mol. Biol. and Evol. \textbf{35}(2), 269--286 (2017)

\bibitem{carter2018hierarchical}
Carter, C.W., Wills, P.R.: Hierarchical groove discrimination by {C}lass {I}
  and {II} aminoacyl-t{RNA} synthetases reveals a palimpsest of the operational
  rna code in the t{RNA} acceptor-stem bases.
\newblock Nucleic Acids Research \textbf{46}(18), 9667--9683 (2018)

\bibitem{carter2015trna}
Carter, C.W., Wolfenden, R.: t{RNA} acceptor stem and anticodon bases form
  independent codes related to protein folding.
\newblock Proc. Natl. Acad. Sci. USA \textbf{112}(24), 7489--7494 (2015)

\bibitem{dayhoff197822}
Dayhoff, M., Schwartz, R., Orcutt, B.: A model of evolutionary change in
  proteins.
\newblock Atlas of Protein Sequence and Structure pp. 345--352 (1978)

\bibitem{delarue2007asymmetric}
Delarue, M.: An asymmetric underlying rule in the assignment of codons:
  possible clue to a quick early evolution of the genetic code via successive
  binary choices.
\newblock RNA \textbf{13}(2), 161--169 (2007)

\bibitem{DiGiulio1995}
Di~Giulio, M.: Was it an ancient gene codifying for a hairpin rna that, by
  means of direct duplication, gave rise to the primitive trna mmolecule?
\newblock J. Theor. Biol. \textbf{177}, 95--101 (1995)

\bibitem{di2001origin}
Di~Giulio, M.: The origin of the genetic code cannot be studied using
  measurements based on the {PAM} matrix because this matrix reflects the code
  itself, making any such analyses tautologous.
\newblock J.Theor. Biol. \textbf{208}(2), 141--144 (2001)

\bibitem{DiGiulio2004}
Di~Giulio, M.: The origin of the t{RNA} molecule: implications for the origin
  of protein synthesis.
\newblock J. Theor. Biol. \textbf{226}, 89--93 (2004)

\bibitem{Wagner1}
Draghi, J., Wagner, G.P.: Evolution of evolvability in a developmental model.
\newblock Evolution \textbf{62}, 301--315 (2007)

\bibitem{Wagner2}
Draghi, J., Wagner, G.P.: The evolutionary dynamics of evolvability in a gene
  network model.
\newblock J. Evol. Biol. \textbf{22}, 599--611 (2009)

\bibitem{eigen1971selforganization}
Eigen, M.: Selforganization of matter and the evolution of biological
  macromolecules.
\newblock Naturwissenschaften \textbf{58}(10), 465--523 (1971)

\bibitem{Facchiano2018}
Facchiano, A., Di~Giulio, M.: The genetic code is not an optimal code in a
  model taking into account both the biosynthetic relationships between amino
  acids and their physicochemical properties.
\newblock J. Theor. Biol. \textbf{459}, 45--51 (2018)

\bibitem{fournier2015ancestral}
Fournier, G., Alm, E.: Ancestral reconstruction of a pre-{LUCA}
  aminoacyl-t{RNA} synthetase ancestor supports the late addition of {T}rp to
  the genetic code.
\newblock J. Mol. Evol. \textbf{80}(3-4), 171--185 (2015)

\bibitem{fuchslin2001evolutionary}
F{\"u}chslin, R.M., McCaskill, J.S.: Evolutionary self-organization of
  cell-free genetic coding.
\newblock Proc. Natl. Acad. Sci. USA \textbf{98}(16), 9185--9190 (2001)

\bibitem{ecodist}
Goslee, S.C., Urban, D.L.: The ecodist package for dissimilarity-based analysis
  of ecological data.
\newblock Journal of Statistical Software \textbf{22}, 1--19 (2007)

\bibitem{Grantham}
Grantham, R.: Amino acid difference formula to help explain protein evolution.
\newblock Science \textbf{185}, 862--864 (1974)

\bibitem{haig1991quantitative}
Haig, D., Hurst, L.D.: A quantitative measure of error minimization in the
  genetic code.
\newblock J. Mol. Evol. \textbf{33}(5), 412--417 (1991)

\bibitem{Hamilton}
Hamilton, W.D.: Narrow roads of gene land: The collected papers of {W}. {D}.
  {H}amilton volume 1: Evolution of social behaviour  (1996)

\bibitem{henikoff1992amino}
Henikoff, S., Henikoff, J.G.: Amino acid substitution matrices from protein
  blocks.
\newblock Proc. Natl. Acad. Sci. USA \textbf{89}(22), 10915--10919 (1992)

\bibitem{hornoshornos}
Hornos, J.E.M., Hornos, Y.M.: Algebraic model for the evolution of the genetic
  code.
\newblock Phys. Rev. Lett. \textbf{71}(26), 4401 (1993)

\bibitem{ikehara2005possible}
Ikehara, K.: Possible steps to the emergence of life: The {GADV}-protein world
  hypothesis.
\newblock The Chemical Record \textbf{5}(2), 107--118 (2005)

\bibitem{jones1992rapid}
Jones, D.T., Taylor, W.R., Thornton, J.M.: The rapid generation of mutation
  data matrices from protein sequences.
\newblock Bioinformatics \textbf{8}(3), 275--282 (1992)

\bibitem{kaiser2018backbone}
Kaiser, F., Bittrich, S., Salentin, S., Leberecht, C., Haupt, V.J., Krautwurst,
  S., Schroeder, M., Labudde, D.: Backbone brackets and arginine tweezers
  delineate class {I} and class {II} aminoacyl t{RNA} synthetases.
\newblock PLoS Comp. Biol. \textbf{14}(4), e1006101 (2018)

\bibitem{koonin2009origin}
Koonin, E.V., Novozhilov, A.S.: Origin and evolution of the genetic code: the
  universal enigma.
\newblock IUBMB Life \textbf{61}(2), 99--111 (2009)

\bibitem{le2008improved}
Le, S.Q., Gascuel, O.: An improved general amino acid replacement matrix.
\newblock Mol. Biol. Evol. \textbf{25}(7), 1307--1320 (2008)

\bibitem{li2013aminoacylating}
Li, L., Francklyn, C., Carter, C.W.: Aminoacylating urzymes challenge the {RNA}
  world hypothesis.
\newblock J. Biol. Chem. \textbf{288}(37), 26856--26863 (2013)

\bibitem{Niefind1991}
Niefind, K., Schomburg, D.: Amino acid similarity coefficients for protein
  modeling and sequence alignment derived from main-chain folding angles.
\newblock J. Mol. Biol. \textbf{219}, 481--497 (1991)

\bibitem{o2003evolution}
O'Donoghue, P., Luthey-Schulten, Z.: On the evolution of structure in
  aminoacyl-t{RNA} synthetases.
\newblock Micro. Mol. Biol. Rev. \textbf{67}(4), 550--573 (2003)

\bibitem{pokarowski2007ideal}
Pokarowski, P., Kloczkowski, A., Nowakowski, S., Pokarowska, M., Jernigan,
  R.L., Kolinski, A.: Ideal amino acid exchange forms for approximating
  substitution matrices.
\newblock Proteins: Structure, Function, and Bioinformatics \textbf{69}(2),
  379--393 (2007)

\bibitem{Popinga}
Popinga, A., Carter, C.W., Bouckaert, R., Wills, P.R.: {Structure-informed
  phylogenetic analysis of the aminoacyl-tRNA synthetases. In: Popinga, A.:
  From the origins of life to epidemics: Bayesian inference, simulation, and
  dynamics of bioinformatic systems}.
\newblock PhD Thesis, Computer Science, University of Auckland: Supplementary
  Data  (2019).
\newblock {http://github.com/alexpopinga/aaRS-Pipeline, accessed 11 April
  2019.}

\bibitem{Rprogram}
{R Core Team}: R: A Language and Environment for Statistical Computing.
\newblock R Foundation for Statistical Computing, Vienna, Austria (2013).
\newblock \urlprefix\url{http://www.R-project.org/}

\bibitem{richards1977areas}
Richards, F.M.: Areas, volumes, packing, and protein structure.
\newblock Annual review of biophysics and bioengineering \textbf{6}(1),
  151--176 (1977)

\bibitem{RodinOhno1995}
Rodin, S.N., Ohno, S.: Two types of aminoacyl-t{RNA} synthetases could be
  originally encoded by complementary strands of nucleic acids.
\newblock Orig. Life Evol. Biosph. \textbf{25}, 565--589 (1995)

\bibitem{schimmel1993operational}
Schimmel, P., Giege, R., Moras, D., Yokoyama, S.: An operational {RNA} code for
  amino acids and possible relationship to genetic code.
\newblock Proc. Natl. Acad. Sci. USA \textbf{90}(19), 8763--8768 (1993)

\bibitem{SmithHartman2015}
Smith, T.F., Hartman, H.: The evolution of class {II} aminoacyl-t{RNA}
  synthetases and the first code.
\newblock FEBS Lett. \textbf{589}, 3499--3507 (2015)

\bibitem{vstambuk2016miyazawa}
{\v{S}}tambuk, N., Konjevoda, P., Manojlovi{\'c}, Z.: Miyazawa-jernigan contact
  potentials and {C}arter-{W}olfenden vapor-to-cyclohexane and
  water-to-cyclohexane scales as parameters for calculating amino acid pair
  distances.
\newblock In: International Conference on Bioinformatics and Biomedical
  Engineering, pp. 358--365. Springer (2016)

\bibitem{StephensonFreeland}
Stephenson, J.D., Freeland, S.J.: Unearthing the root of amino acid similarity.
\newblock J. Mol. Evol. \textbf{77}(4), 159--169 (2013)

\bibitem{Turing}
Turing, A.: The chemical basis of morphogenesis.
\newblock Philos. Trans. R. Soc. B (London) \textbf{237}, 37--72 (1952)

\bibitem{Unvert2017}
Unvert, K.E., Kovacs, F.A., Zhang, C., Hellmann-Whitakerc, R.A., Arndt, K.N.:
  Evolution of leucyl-t{RNA} synthetase through eukaryotic speciation.
\newblock American Journal of Undergraduate Research \textbf{14}, 69--83 (2017)

\bibitem{vetsigian2006collective}
Vetsigian, K., Woese, C., Goldenfeld, N.: Collective evolution and the genetic
  code.
\newblock Proc. Natl. Acad. Sci. USA \textbf{103}(28), 10696--10701 (2006)

\bibitem{whelan2001general}
Whelan, S., Goldman, N.: A general empirical model of protein evolution derived
  from multiple protein families using a maximum-likelihood approach.
\newblock Mol. Biol. Evol. \textbf{18}(5), 691--699 (2001)

\bibitem{wills1993self}
Wills, P.R.: Self-organization of genetic coding.
\newblock J. Theor. Biol. \textbf{162}(3), 267--287 (1993)

\bibitem{WillsMutualOrdering}
Wills, P.R.: Spontaneous mutual ordering of nucleic acids and proteins.
\newblock Orig. Life Evol. Biosph. \textbf{44}, 293--298 (2014)

\bibitem{wills2018insuperable}
Wills, P.R., Carter, C.W.: Insuperable problems of the genetic code initially
  emerging in an {RNA} world.
\newblock Biosystems \textbf{164}, 155--166 (2018)

\bibitem{wills2015emergence}
Wills, P.R., Nieselt, K., McCaskill, J.S.: Emergence of coding and its
  specificity as a physico-informatic problem.
\newblock Orig. Life Evol. Biosph. \textbf{45}(1-2), 249--255 (2015)

\bibitem{woese1965evolution}
Woese, C.R.: On the evolution of the genetic code.
\newblock Proc. Natl. Acad. Sci. USA \textbf{54}(6), 1546--1552 (1965)

\bibitem{wolf2007origin}
Wolf, Y.I., Koonin, E.V.: On the origin of the translation system and the
  genetic code in the {RNA} world by means of natural selection, exaptation,
  and subfunctionalization.
\newblock Biology Direct \textbf{2}(1), 14 (2007)

\bibitem{WolfendenCarter2015}
Wolfenden, R., Lewis, C.A., Yuan, Y., Carter, C.W.: Temperature dependence of
  amino acid hydrophobicities.
\newblock Proc. Nat. Acad. Sci. USA \textbf{112}, 7484--7488 (2015)

\bibitem{wong1975co}
Wong, J.T.F.: A co-evolution theory of the genetic code.
\newblock Proc. Natl. Acad. Sci. USA \textbf{72}(5), 1909 (1975)

\bibitem{Yampolsky2005}
Yampolsky, L.Y., Stoltzfus, A.: The exchangeability of amino acids in proteins.
\newblock Genetics \textbf{170}, 1459--1472 (2005)

\bibitem{Wolf1999}
Yuri I.~Wolf L.~Aravind, N.V.G., Koonin, E.V.: Evolution of aminoacyl-t{RNA}
  synthetases—analysis of unique domain architectures and phylogenetic trees
  reveals a complex history of horizontal gene transfer events.
\newblock Genome Research \textbf{9}, 689--710 (1999)

\bibitem{Zimmerman}
Zimmermann, K., Gibrat, J.: Amino acid ``little {B}ig {B}ang'': Representing
  amino acid substitution matrices as dot products of {E}uclidian vectors.
\newblock BMC Bioinformatics \textbf{11:4} (2010)

\end{thebibliography}

%
%

\end{document}